\documentclass[10pt,letterpaper]{article}
\usepackage[top=0.85in,left=2.75in,footskip=0.75in]{geometry}

\usepackage{amsmath,amssymb}

\usepackage{changepage}

\usepackage[utf8x]{inputenc}

\usepackage{textcomp,marvosym}

\usepackage{cite}

\usepackage{nameref,hyperref}

\usepackage[right]{lineno}

\usepackage{microtype}
\DisableLigatures[f]{encoding = *, family = * }

\usepackage[table]{xcolor}

\usepackage{array}

\newcolumntype{+}{!{\vrule width 2pt}}

\newlength\savedwidth

\raggedright
\usepackage[aboveskip=1pt,labelfont=bf,labelsep=period,justification=raggedright,singlelinecheck=off]{caption}

\makeatletter
\renewcommand{\@biblabel}[1]{\quad#1.}
\makeatother

\date{}

\usepackage{lastpage,fancyhdr,graphicx}
\usepackage{epstopdf}
\fancyheadoffset[L]{2.25in}
\fancyfootoffset[L]{2.25in}

\newcommand{\phit}{P_{hit}}
\newcommand{\beginsupplement}{
        \setcounter{table}{0}
        \renewcommand{\thetable}{S\arabic{table}}
        \setcounter{figure}{0}
        \renewcommand{\thefigure}{S\arabic{figure}}
     }

\begin{document}
\vspace*{0.2in}

\begin{flushleft}
{\Large
\textbf\newline{Collective Decision Dynamics in Group Evacuation: Modeling Tradeoffs and Optimal Behavior} 
}
\newline
\\
Kimberly J. Schlesinger\textsuperscript{1*},
Chantal Nguyen\textsuperscript{1},
Imtiaz Ali\textsuperscript{1,2},
Jean M. Carlson\textsuperscript{1}
\\
\bigskip
\textbf{1} Department of Physics, University of California, Santa Barbara, CA, USA
\\
\textbf{2} Department of Physics, University of California, Merced, CA, USA
\\
\bigskip

* kschlesi@physics.ucsb.edu

\end{flushleft}
\section*{Abstract}
Quantifying uncertainties in collective human behavior and decision making is crucial for ensuring public health and safety, enabling effective disaster response, informing the design of transportation and communication networks, and guiding the development of new technologies. However, modeling and predicting such behavior is notoriously difficult, due to the influence of a variety of complex factors such as the availability and uncertainty of information, the interaction and influence of social groups and networks, the degree of risk or time pressure involved in a situation, and differences in individual personalities and preferences. Here, we develop a stochastic model of human decision making to describe the empirical behavior of subjects in a controlled experiment simulating a natural disaster scenario. We compare the observed behavior to that of statistically optimal Bayesian decision makers, quantifying the extent to which human decisions are optimal and identifying the conditions in which sub-optimal decisions are made. Finally, we investigate how human evacuation strategies change when decisions are made in groups under a variety of different rules, and whether these group strategy adjustments are optimal or beneficial.



\section*{Introduction}
Collective human decision making is central to the behavior of social and political systems, and plays a crucial role in transportation and communication networks. The ability to characterize and predict collective decision dynamics is thus essential for the design and control of many types of networked systems, and for the creation of effective strategies for interacting with these systems, especially in scenarios such as network failures or large-scale natural disasters. However, accurately modeling collective human behavior remains a significant challenge due to the complexity of the problem. Decisions are influenced by an interplay of multiple factors, such as the spread and reliability of information, time pressure, perceived risks, social interactions, and differences between individuals~\cite{stir7, stir8, evachouse, stir10}. Moreover, human factors themselves are a primary source of fragility in communication, transportation, and other networked systems upon which these dynamics play out~\cite{huang2011,meshkati1991}. 

In this paper, we develop a model for collective decision dynamics that isolates key factors and tradeoffs affecting individual decisions, and quantifies their impact on the dynamics of the population. We apply this model to data collected in a controlled experiment that simulates decision making about whether and when to evacuate in the face of an impending natural disaster. The simplicity and flexibility of our modeling approach enables the quantification the influence of specific variables on evacuation behavior, such as the reported likelihood of a disaster strike and the protocols governing how decisions are made. With this model, we probe and quantitatively characterize several complex situations that affect collective dynamics, including forced group-consensus decisions, competition for space in an evacuation shelter, the learning of strategies over time, and the contrast between observed and optimal behavior.

The experimental data used in these models was first presented in~\cite{fangqiu}, and the experimental design is based upon that in~\cite{sean}, in which a virtual natural disaster scenario is simulated in a controlled laboratory setting, and participants must decide whether or not to evacuate based on the information they are given about the disaster and the actions of their neighbors. This controlled setup approximates many of the factors that influence decisions in real natural disasters, including time and social pressure, competition for resources, making uncertain choices under risk of loss, and group consensus decision strategies. Moreover, by systematically varying experimental parameters and gathering precise data on participants' actions and decisions, we can build and test a model of the factors that most influence group decision dynamics. Our modeling approach in this work, combined with this unique experimental design, allows the systematic examination of specific factors that cannot be tested and controlled for in surveys of population-wide behavior, while maintaining a context that can inform models of evacuation behavior in real disaster scenarios. The resulting models capture important tradeoffs, and can be used to incorporate the effects of these human factors into more complex and large-scale models of collective human phenomena.

\textbf{Background.}
Complexity is a fundamental challenge in modeling collective behavior, due to both the influence of multiple factors on human decisions, and the influence that human factors themselves have on the breakdown or failure of networked systems that support them. This is often compounded by the added difficulty of accurate measurement, especially in situations involving risk and danger, such as the human response to a natural disaster. When the first priorities are emergency response and perhaps evacuation in the face of a wildfire or a hurricane, there are often insufficient time and resources to support collecting complete data on the movements, communications, and decisions of the affected population. However, these threatening and uncertain scenarios often display noisy and fragile individual performance that lead to non-optimal collective outcomes~\cite{evachouse, stir10}, and common assumptions of either optimal or random individual behavior do not reliably hold. Measuring factors that shape these individual decisions, and incorporating them into accurate and predictive models of behavior, is crucial for designing evacuation and communication strategies that can anticipate and avoid these breakdowns.

Current efforts to simulate the large-scale evacuation of populations often rely on data from surveys in which participants provide the details of their evacuation decisions and movements after a disaster has passed~\cite{gladwinpeacock,katrina1,katrina2}. This provides a basis from which to examine behavior at a population scale, although it leads to extremely complex models with many factors and parameters, and is prone to biases and errors in self-reporting. In contrast, many psychology and neuroscience studies probe and model the act of decision-making itself in abstract problems or in closely controlled laboratory experiments~\cite{beck2008,leonard1}. This provides precise measurements of neural or behavioral aspects of decision making under certain conditions, but application of these results to disaster-like scenarios that involve extensive social and environmental influences is difficult to implement. Here, we demonstrate an approach that uses a controlled experiment to measure key factors driving evacuation decision dynamics, and provides a quantitative model suitable for incorporating these insights into larger-scale simulations.

In this work we focus specifically on social and group decision making. Human and animal populations constantly use information about their social contacts' behavior and beliefs to guide their own decisions~\cite{mavrodiev2013,dani,nedic2012,fishgroups}, and these influences can factor into decision making in several distinct ways.
First, social pressures and social communication patterns have the ability to influence individual decisions -- a single actor may change the content or timing of a decision based upon the opinions or decisions of others, or based on other information accessed through their social network, whether accurate or inaccurate. This involves both weighing the influence of others' actions on one's own social position or probability of success, as well as considering the effect of one's own individual decision upon the group.

In some situations, the tie between individual decisions and group dynamics becomes even more explicit when decisions are constrained to be made as a group consensus, following a certain imposed protocol. Common examples of this type of scenario include governmental procedures in which citizens use a majority vote to choose representatives; teams in business environments with a designated leader and rules to facilitate discussion and produce consensus; and military squadrons, whose strategy hinges upon following the decisions of an appointed commander. In these cases, group members must all abide by the consensus decision, although there are many possible protocols for making this decision that give different amounts of influence to each group member, depending on the situation and the goals of the group. It is not well understood how the success of a group in a given situation is affected by the chosen decision protocol, as well as by the composition of the group and the differences between the individuals within it, but insights into optimal group actions can help inform and improve the decisions made across such scenarios, from military training to presidential elections.

In natural disaster scenarios in particular, group decision making has a significant influence on the dynamics. Evacuations are often made in groups, such as households, who may be bound by a common mode of transportation or shared resources, but have different strategies for coming to a decision about whether and when to evacuate~\cite{evachouse,huang2011,heath2001human}. Thus, the decision dynamics of the population may fundamentally change compared to what a model of individually-driven behavior would predict. Groups can exhibit a tendency to make riskier collective decisions after group discussion than individuals would alone~\cite{stoner, isenberg}, and social interactions can lead to a ``mob mentality"~\cite{stir12, stir13, stir14} that may heavily influence evacuation decisions~\cite{stir7, stir8}. In this work, in order to examine the dynamics of group evacuation decisions, we collect information both on the participants' decisions as individuals, and on their decisions when constrained to evacuate in groups under various decision protocols. Understanding how to predict the activity of a population acting upon group decisions is crucial to the design of effective evacuation policies. 

Many studies have investigated whether humans make optimal decisions under various conditions, weighing their evidence for a decision and the uncertainties they face in order to make the choice with the best possible expected outcome~\cite{stir5, kearns1, leonard1, beck2008}. This question is relevant in many decision making scenarios, from individual choices to collective behavior, and important to consider in designing strategies for interacting with and influencing such systems. If an optimal approach to an evacuation scenario can be devised, and if specific factors contributing to non-optimal behavior can be identified, then system interventions can be proposed and tested in order to avoid or correct this  behavior, facilitating safer and more efficient responses. In addition, a measure of optimality provides a useful metric against which to evaluate the performance of existing systems and compare various evacuation protocols or strategies.

\textbf{Our approach.}
This modeling work is based upon data from a controlled evacuation experiment, originally presented in Nguyen et al.~\cite{fangqiu}. While~\cite{fangqiu} discusses the experimental design in detail and presents statistical analyses of the results, this work pursues a distinct and complementary decision modeling approach using the collected evacuation data. 

Nguyen et al.~\cite{fangqiu} introduces a machine learning neural network model, which serves as a data-driven tool to predict the precise evacuation times of participants or groups based on factors such as group decision method, group size, the likelihood of the disaster striking and how rapidly that likelihood is changing over time, and personal identifying characteristics of each participant. The neural network predicts precise evacuation times with an accuracy rate of 85\%, indicating that the information contained in these factors is sufficient to accurately determine evacuation behavior. By using experimental observations alone and training on thousands of samples of actual decisions and their contexts, the neural network identifies correlation structure in the data in an unbiased manner, and highlights how the addition of new types of information improves the prediction accuracy of the decision model.

This work takes an alternative approach with fundamentally different aim. Here, we build an analytical decision model that isolates a few especially important factors and tradeoffs, provides a predictive description of how these factors influence evacuation behaviors, and quantifies the difference between individual evacuation outcomes, group outcomes, and optimal outcomes. While the neural network model of Nguyen et al.~\cite{fangqiu} is a valuable and unbiased method for determining how well the information in a set of factors can determine evacuation outcomes, the model introduced in this paper provides a more complete and systematic approach to critical modeling questions concerning the key mechanisms behind evacuation decisions, into which a neural network can give only limited insight. This 
advances the fundamental understanding of underlying human factors which may enhance or degrade network performance in numerous applications, and the resulting insights may be incorporated into a large-scale simulations of natural disasters and other collective decision making scenarios.

The decision model introduced here provides a framework for pursuing quantitative questions about group and optimal decision making. Starting from a relatively simple two-parameter Markov model for evacuation decision making over the course of a natural disaster, inspired by~\cite{sean}, we first demonstrate that the model can be fit to accurately predict the population's behavior when decisions are made individually. Then we extend the model to quantitatively evaluate the differences between individual and group decision making, investigating to what extent individual strategies are altered when acting in a group scenario, and how the size and decision mechanism of a group influences its overall performance. We also use Bayesian parameter inference to approximate the best possible strategies for participants in the disaster experiment, and determine quantitatively whether human decision makers act optimally in the face of an uncertain and potentially dangerous threat. In particular, we ask whether non-optimal evacuation decisions are influenced by identifiable factors, which could inform better evacuation planning and help to optimize disaster response protocols.

We develop and test these methods in the case of predicting the evacuation behavior of human populations in a natural disaster scenario. However, the framework presented here is flexible enough to identify and capture key tradeoffs in a broader class of systems that involve collective decision dynamics, allowing for more accurate representation of human factors in complex computational modeling, policy, and management of real world disaster situations.

\section*{Materials and Methods}
\textbf{Behavioral experiment.} This work addresses the question of human decision making in an evacuation scenario by examining data from a behavioral experiment performed on March 13, 2015, at the University of California, Santa Barbara (UCSB). This experiment was designed to simulate certain aspects of a natural disaster scenario in a controlled setting, while collecting detailed information on the behavior and evacuation decisions of the participants.

In this experiment, a virtual community of 50 participants (37 male, 7 female, 6 not specified or other) decided if and when to evacuate from a potential natural disaster in a series of 144 trials, each lasting up to 30 seconds. The experiment was approved by the Institutional Review Board, Office of Research, UCSB, protocol number 15-0010. All participants provided written informed consent and were paid for their participation.

A complete description of the experimental protocol can be found in~\cite{fangqiu}. The present work builds a quantitative and flexible decision model using empirical data and statistical insights from this behavioral experiment. In this methods section, we present a summary of the experimental protocol which discusses points relevant to our current modeling approach; full details are given in~\cite{fangqiu}, along with a presentation of the results and statistical analysis.

\begin{figure}[h!]
\includegraphics[width=\linewidth]{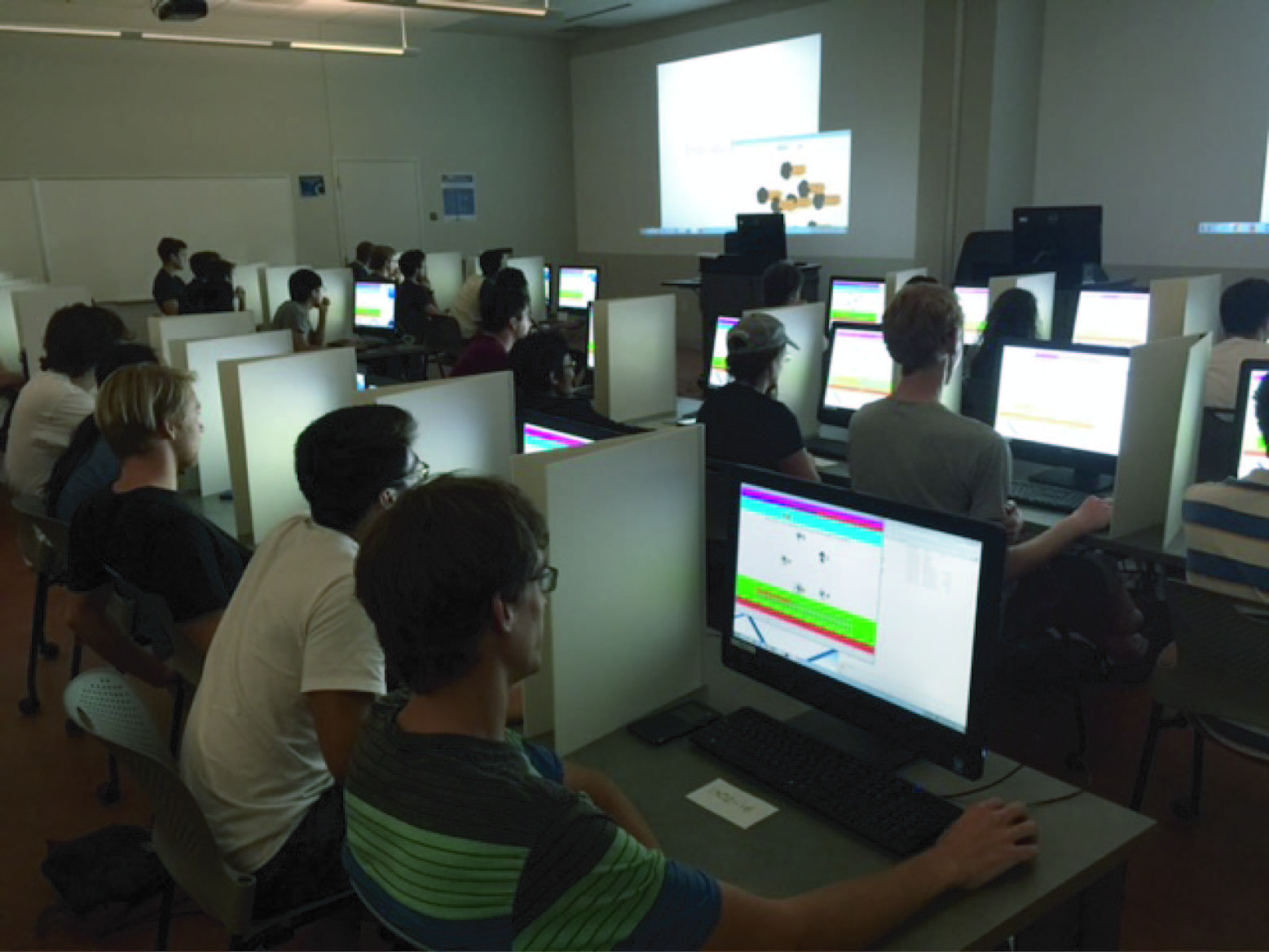}
\caption{ {\bf Experimental setup.} 50 participants made evacuation decisions in simulated natural disaster scenarios, based on information provided via personal computer interfaces.}
\label{fig:exp}
\end{figure}

\begin{figure}[h!]
\includegraphics[width=\linewidth]{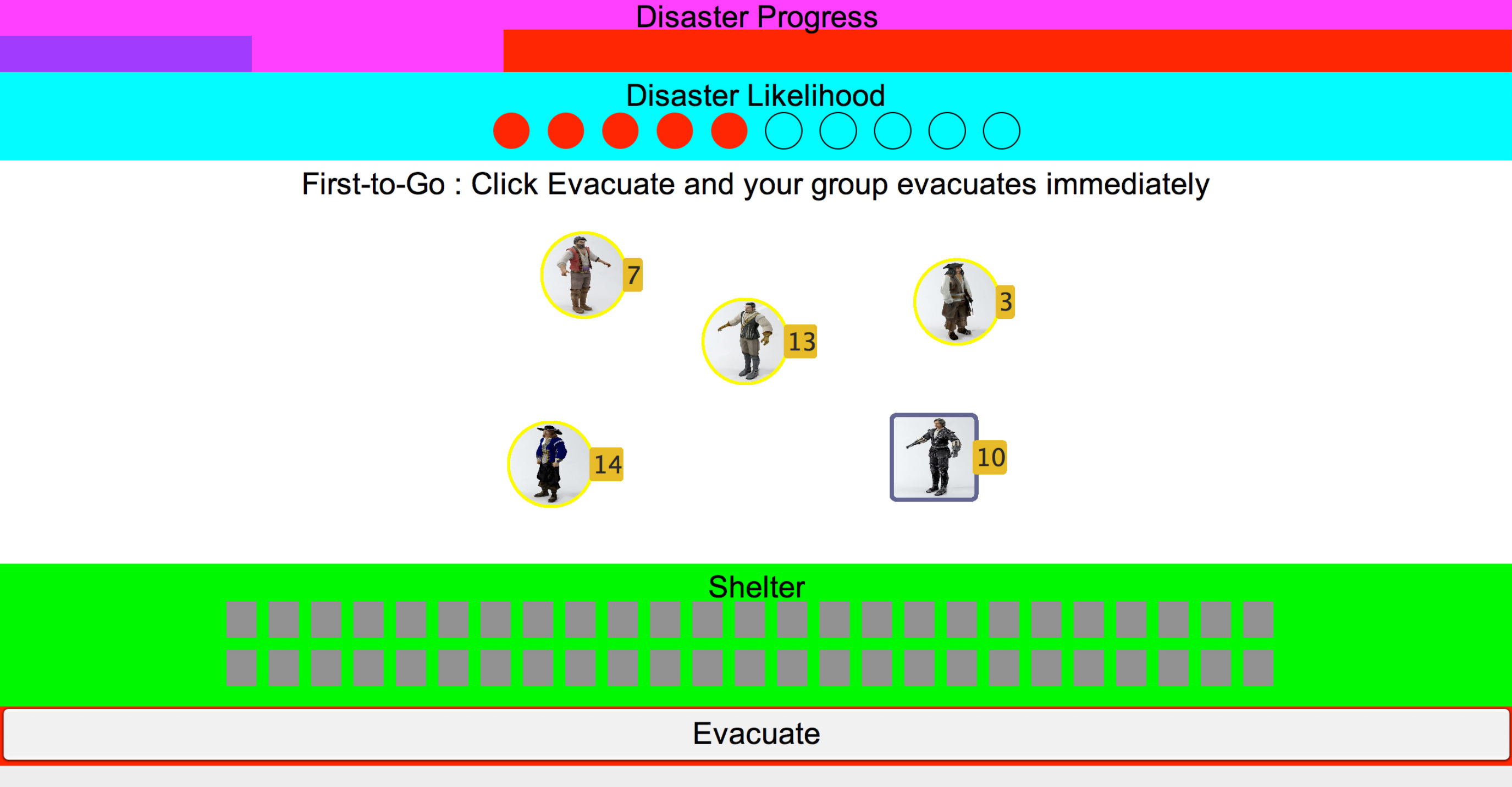}
\caption{{\bf Experiment interface.} An example of the experiment interface used by participants. Disaster progress is shown by a blue bar progressing with time along the top of the screen; once the blue bar enters the red region, the disaster may strike at any time. The likelihood of the disaster striking is depicted by the red circles, which are updated on each time step; here, five out of ten circles are filled in, for a likelihood of 50\%. This example is from a group trial, where the group decision protocol is explained, and the participant can see a depiction of the other members of their group and their score ranks. The beds in the shelter are also depicted; this trial has 50 available beds, none of which have been filled so far. At the bottom of the screen is the evacuate button, which the participant clicks in order to evacuate.}
\label{fig:exp_sshot}
\end{figure}

Participants in the experiment used individual computer interfaces (Figs.~\ref{fig:exp} and \ref{fig:exp_sshot}), where they were presented with information regarding the likelihood of disaster strike, denoted $\phit$, as well as the availability of spaces in an evacuation shelter and the behavior of other participants. Information about the progression of the disaster was updated and broadcast at every half-second time step. At the beginning of each trial, participants were in the ``at home'' state; once a participant evacuated, they immediately moved to the ``in shelter'' state and could not return home.

The $\phit$ value varied non-monotonically with time in stochastically-varying step sizes between $0$ and $1$, where $0$ represents a disaster miss, and $1$ represents a disaster strike. The $\phit$ values broadcast to participants were rounded down to the nearest tenth and displayed as red circles that were updated at every time step (see Fig.~\ref{fig:exp_sshot}). During the first 20 time steps of the trial, the disaster was guaranteed to neither strike nor miss, but in the remaining time steps -- up to a total of 60 -- the disaster could strike or miss at any time, which would cause the trial to end.

In each trial, the maximum shelter capacity was either 5, 25, or 50 spaces. As participants evacuated and shelter spaces filled, the remaining available shelter capacity was shown to all participants on their computer interfaces (see Fig.~\ref{fig:exp_sshot}), through a shelter space graphic which was also updated at every time step. 

Participants clicked an ``Evacuate'' button upon deciding to evacuate. Each trial was classified as either an ``individual'' or a ``group'' trial. In individual trials, a decision to evacuate immediately resulted in an evacuation if there was available shelter space. However, in group trials, deciding to evacuate did not necessarily result in an evacuation; a decision to evacuate indicated a vote for collective group evacuation, but the group would not evacuate until enough votes had been accumulated according to an assigned group decision protocol. In these group trials, participants were randomly assigned to groups of 5 or 25, and the groups collectively acted according to one of three protocols: first-to-go (FTG), last-to-go (LTG), and majority-vote (MV). In the first-to-go protocol, once one member of the group decided to evacuate, the entire group would evacuate as well. In the last-to-go protocol, the group would not evacuate until every member of the group decided to evacuate. In the majority-vote protocol, the group would evacuate once a majority of members had decided to evacuate. In group trials, participants were shown the group protocol and their fellow group members, represented by unique avatars which would change shape once the corresponding individual made a decision to evacuate. 

Before each trial, participants were each staked 10 points, out of which a number of points was deducted at the end of the trial based on the success of their decision. The possible values are indicated in a loss matrix (Table \ref{table:lossmat}) which was displayed to participants at the beginning of the experiment, although its importance was not emphasized in order to reduce its effect on decision making. The participants were also ranked by their scores and informed of their rankings after each trial. During group trials, each participant was also shown the rankings of all members of their group. At the end of the experiment, the participants received monetary compensation based on their final rankings.

\begin{table}[ht!]
\centering
\caption{
{\bf Loss matrix.} Number of points deducted and net points gained for each possible outcome of a single trial.}
\begin{tabular} {|l|r|r|}
\hline
\multicolumn{3}{|c|}{\textbf{Points deducted}}\\
\hline
 & Disaster hits & Disaster misses  \\ \hline
At home  & -10 & 0 \\ \hline
In shelter  & -6 & -2 \\
\hline
\multicolumn{3}{|c|}{\textbf{Net points gained}}\\
\hline
 & Disaster hits & Disaster misses  \\ \hline
At home  & 0 & 10 \\ \hline
In shelter  & 4 & 8 \\
\hline
\end{tabular}
\begin{flushleft}
\end{flushleft}
\label{table:lossmat}
\end{table}

In this paper, we denote individual trials by ``Ind50'', ``Ind25'', or ``Ind5'', where the numbers represent the initial shelter capacity on that trial. Group trials are denoted by the group size, i.e. ``group-5'' and ``group-25'' for trials with groups of 5 and groups of 25, respectively, with ``FTG'', ``LTG'', and ``MV'' denoting the protocols of first-to-go, last-to-go, and majority-vote, respectively.

The data from 16 out of 144 trials were excluded due to technical difficulties. A total of 128 trials are included in the analysis. Out of these 128 trials, 46 trials were individual trials (16 Ind50, 13 Ind25, 5 Ind5), and 82 were group trials.

\textbf{Markov model of decision dynamics.}
We use a stochastic Markov model of decision dynamics to fit the observed data. We define the state of the system as an $(N+1)$-element vector $\Psi(t)$. Here, $N=50$ is the total number of participants in the community, and each system state element $\Psi_n(t)$, where $n=0,1,...,N$, gives the probability that a total of $n$ participants have evacuated at time $t$. Since participants are forbidden from returning to their homes once evacuated, the total number $n$ evacuated at any given time $t^\prime$ will necessarily be the sum of all those who evacuate during time step $t^\prime$ and during each of the previous time steps $t=0,1,...,t^\prime-1$. In other words, the vector $\Psi(t)$ describes the probabilities of cumulative evacuation numbers since the beginning of the disaster threat at $t=0$. As a probability vector, it is required that $\sum_n \Psi_n(t) = 1 ~ \forall ~ t$.

We model the time evolution of $\Psi(t)$ as a Markov process driven by a generator matrix $\textbf{A}(t)$:
\begin{equation}\label{eq:mastereq}
\frac{d\Psi(t)}{dt} = \textbf{A}(t)\Psi(t).
\end{equation}
Each entry $A_{ij}(t)$ gives the expected transition rate from state $S_j$ to $S_i$ at time $t$. That is, if $p[n,t] = S_n(t)$ gives the probability of the system being in state $n$ at time $t$, then $A_{ij}(t) = p[j,t+\delta t|i,t]$. Since individuals cannot return home once evacuated, the total number $n$ of evacuated individuals must either increase or stay constant over time; thus, the probability of a transition from state $i$ to $j$, where $j<i$, is equal to zero and $\textbf{A}$ is a lower-triangular matrix.

For simplicity, we treat the decisions of the individual participants as independent and identically distributed. Thus, at a given time, the probability of participants to evacuate is modeled as a binomial distribution (sum of Bernoulli distributions), such that the probability that exactly $j-i$ individuals evacuate in $[t,t+\delta t]$ is given by
\begin{equation}
p[j,t+\delta t|i,t] = A_{ij}(t) = \binom{N-i}{j-i} q(t)^{j-i}(1-q(t))^{N-j}.
\end{equation}
Here, $q$ describes the \textit{individual decision model}, or the probability that a single individual will evacuate in this time interval, and it may in general be dependent on time, disaster probability, shelter space, and any number of other factors. We begin by describing $q$ as a simple function of the disaster probability, $\phit(t)$, which is itself time-dependent over the course of a disaster scenario. We choose a two-parameter, monotonically increasing power-law function:
\begin{equation}\label{eq:decmodel}
q(\phit) = a \left(\phit\right)^b,
\end{equation}
where $a$ is the maximum value of $q$ -- i.e., the probability of evacuation when $\phit(t) = 1$ -- and $b$ defines the shape of the function. $b = 1$ gives a linear dependence, while $b > 1$ gives a concave function approaching a step function as $b \rightarrow \infty$. ($0 < b < 1$ gives a convex function which does not accurately describe the behavior observed.)

\textbf{Fitting to empirical data.}
To fit this model to the empirical data, we first define the variables $H_\nu$ and $J_\nu$, where $H_\nu$ gives the total number of times that participants who were \textit{at home} saw the probability $\phit = \nu \in \{0.1, 0.2, \hdots, 0.9, 1.0\}$, summed over all participants and all trials, and where $J_\nu$ gives the total number of times that participants who saw the probability $\phit = \nu$ evacuated on that same time step. 

We can also define the overall \textit{evacuation rate} $\Theta_\nu$ as a stochastic process for each value of $\phit = \nu$. If the distribution of $J_\nu$ given $H_\nu$ and $\Theta_\nu$ is binomial, then the distribution of $\Theta_\nu$ given $H_\nu$ and $J_\nu$ is given by the conjugate prior, the beta distribution, denoted Beta$(\alpha,\beta)$, with parameters $\alpha = J_\nu +1$ and $\beta = H_\nu - J_\nu +1$~\cite{otto2007}. We then formulate the likelihood of the measured rate $\Theta_\nu$ (or, equivalently, the empirical $H_\nu$ and $J_\nu$), given the model $q(\phit=\nu) \equiv q_\nu$ in Eq.~\eqref{eq:decmodel} and its parameters $a$ and $b$:
\begin{eqnarray}
p[\Theta_\nu|q_\nu(a, b)] & = & \prod_\nu \mathrm{Beta}(J_\nu +1,H_\nu - J_\nu +1) \\
& = & \prod_\nu \frac{1}{B(\alpha,\beta)}q_\nu^{J_\nu}(1-q_\nu)^{H_\nu - J_\nu}, \label{eq:beta}
\end{eqnarray}
where $B(\alpha,\beta)$ denotes the beta function of the beta distribution parameters $\alpha, \beta$.

We then find the parameters $a$, $b$ that maximize the likelihood of the observed $H_\nu$ and $J_\nu$ values by minimizing the following expression, which is proportional to the negative logarithm of this likelihood, with respect to the parameters of $q_\nu$~\cite{sean}:
\begin{equation} \label{eq:ll}
-\sum_\nu [J_\nu~\ln(q_\nu)+(H_\nu - J_\nu)~\ln(1-q_\nu)].
\end{equation}
We obtain standard deviations on these fits of the parameters $a$ and $b$ with a bootstrapping procedure. We fit the model to 1000 artificial data sets generated by random sampling with replacement from the original pool of trials, and calculate the standard deviation over these fits.

For trials with limited shelter capacity, we reformulate the model $q$ to reflect the dependence of the initial shelter capacity $s$ on individual decision making, such that $q$ is a function of both $\phit$ and $s$. Since we expect that for limited shelter space, participants will be more likely to evacuate at lower values of $\phit$, the exponent $b$ should be an increasing function of the shelter space. We set $b = s/c$ such that
\begin{equation} \label{eq:sp}
q(\phit,s) = a \phit^{s/c}.
\end{equation}
Furthermore, to fit $q(\phit,s)$ to the data, we define $H^{\prime}_{\nu}$ to represent the total number of times that participants at home observed $\phit = \nu$ {while the shelter was not full}, since a trial will not end when the shelter reaches full capacity if $\phit$ has not yet reached 0 or 1, but further observed $\phit$ values cannot affect participants' actions once the shelter is full.

\textbf{Evaluating model accuracy.}
In order to evaluate the ability of this model to predict individual evacuation behavior and to ensure it does not overfit to noise or individual variation in the data, we perform leave-one-out cross-validation for each set of individual trials with the same shelter capacity. For each of the trials in a set, we exclude data from one trial, train the model on the remaining trials with the previously described MLE fitting process, and compare the resulting prediction of behavior in the excluded trial with that of the original model. We quantify this comparison by reporting the root mean square error (RMSE) between the LOOCV prediction and the observed behavior, where the mean is over all time steps of the trial.

We extend this procedure to quantify the extent to which this model, when trained on individual decision protocol trials, can predict decision-making behavior in the three group protocols: first-to-go (FTG), last-to-go (LTG), and majority-vote (MV). We term this the ``na\"{i}ve cross-validation'' method of determining the relation between individual and group behavior.

\textbf{Behavioral simulation.}
Once we have determined maximum-likelihood model parameters, we can use them to numerically solve the master equation (Eq.~\eqref{eq:mastereq}), by integrating the differential equations through time from a given initial condition. This provides a full probability distribution over possible system states at each point in time. However, to reveal the nature of the dynamics of individual disaster-response episodes, rather than simply obtaining moments of the full time-evolving distribution, we also simulate single trajectories of the system through stochastic sampling. We use the Gillespie-Doob algorithm to stochastically generate system trajectories with probabilities identical to those defined in the master equation~\cite{gillespie}. These sampled instances of system behavior allow us to reconstruct examples of the decisions of each individual and of the cumulative evacuations in the population. 

We use this simulation procedure to compare individual and group behavior by enforcing group decision making protocols on the sampled evacuation times, which we call the ``grouped individual simulation'' procedure. This simulates the behavior of a population that uses the individual decision strategy of the basic model even when operating under group decision protocols.

\textbf{Determining optimal behavior.}
We use two methods to determine the \textit{optimal} strategy (i.e., set of decision model parameters) to follow in the disaster scenarios of the experiment. First, we make the rather simplistic assumption that the strategy does not evolve with time, either over the course of a single trial or over several trials. We compute the expected value of the final score an individual would obtain on the presented ensemble of evacuation trajectories, given a static set of parameters to characterize the decision model in Eq.~\eqref{eq:decmodel}.

For a single individual $\psi$ in the model, the probability of having evacuated at time $t$ during a trial, or $\psi_{evac}(t)$, evolves as
\begin{equation}\label{eq:indprob}
\frac{d~\psi_{evac}(t)}{dt} = a \phit(t)^b \left[1 - \psi_{evac}(t)\right].
\end{equation}
Clearly, this probability will either increase or remain constant over the course of a trial. The expectation value of the score attained in a trial is based upon this distribution at time $t_{end}$, when the trial concludes (i.e., the time when the disaster probability hits 1 or 0); the trial outcome; and the loss matrix used for calculating scores (which is invariant across trials). The loss matrix used gives the score or loss $S$ for the four possible trial outcomes and is shown in Table \ref{table:lossmat}. 

The total expectation of the score $S$ for a subject $\psi$ on a given trial $g$ is then calculated as
\begin{equation}\label{eq:exp_score_hit}
\langle S^\psi_g \rangle_H = [\psi_{evac}(t^g_{end})S_{H,E} + (1-\psi_{evac}(t^g_{end}))S_{H,NE}]
\end{equation}
if the disaster hit during trial $g$, and
\begin{equation}\label{eq:exp_score_miss}
\langle S^\psi_g \rangle_M = [\psi_{evac}(t^g_{end})S_{M,E} + (1-\psi_{evac}(t^g_{end}))S_{M,NE}]
\end{equation}
if the disaster missed during trial $g$. The total expected score is given by the sum of the expected scores over all trials:
\begin{equation}\label{eq:exp_score_tot}
\langle S^\psi \rangle = \sum_{g} \langle S^\psi_g \rangle_H ~\delta(P_{hit}^g(t^g_{end}),1) + \langle S^\psi_g \rangle_M~\delta(P_{hit}^g(t^g_{end}),0)
\end{equation} 
We optimize this total expected score over a range of parameters to obtain the \textit{static optimal} strategy. 

To create a more realistic and advantageous model of optimal behavior, we also model individuals as optimal Bayesian observers, who update their strategies over time to reflect the parameters that give the highest \textit{a posteriori} expected value of reward. The result is the \textit{Bayesian optimal strategy}, an evolving series of parameter values that change as the player learns more about the trial. Beginning with a uniform prior over a range of parameter space, the Bayesian-optimal player will calculate, after each trial, the likelihood of obtaining the highest possible reward, given the parameters. For trials in which the disaster hits, the maximum score is achieved by evacuating; for trials in which the disaster misses, the maximum score is achieved by remaining at home. Hence, the likelihood of a player achieving the maximum score for trial $g$, given a set of parameters $\theta$, is given by
\begin{equation}
\begin{split}
\mathcal{L}(\text{max score}|\theta)  = &\ \psi_{evac}(t_{end}^g) \delta (\phit^g(t_{end}^g),1) \\ &+ (1-\psi_{evac}(t_{end}^g))\delta (\phit^g(t_{end}^g),0). 
\end{split}
\end{equation}
After each trial, the likelihoods evaluated over the parameter space are used to update the posterior probability of receiving the highest possible total score. Thus, after each trial, the Bayesian-optimal player will adjust behavior to take into account all available evidence for the best strategy, while avoiding a strategy that overfits to the specific noise in a particular disaster trajectory. 

We use these two definitions of optimality to test whether individual decision makers, or decision makers acting under group protocols, act in an optimal manner during the evacuation scenarios.

\section*{Results}
We begin by fitting the model described above to the sixteen Ind50 trials -- those with individual decision protocols (no groups) and shelter space available for all players (i.e., 50 initial spaces). This provides a basic estimate of the model's ability to predict evacuation behavior in the population based only on $\phit$. Subsequently, we extend our model to include factors such as limited shelter space and decision making under group protocols, comparing these results to the simpler model to ascertain the importance of these factors in predicting behavior.

\begin{figure}[h!]
\includegraphics{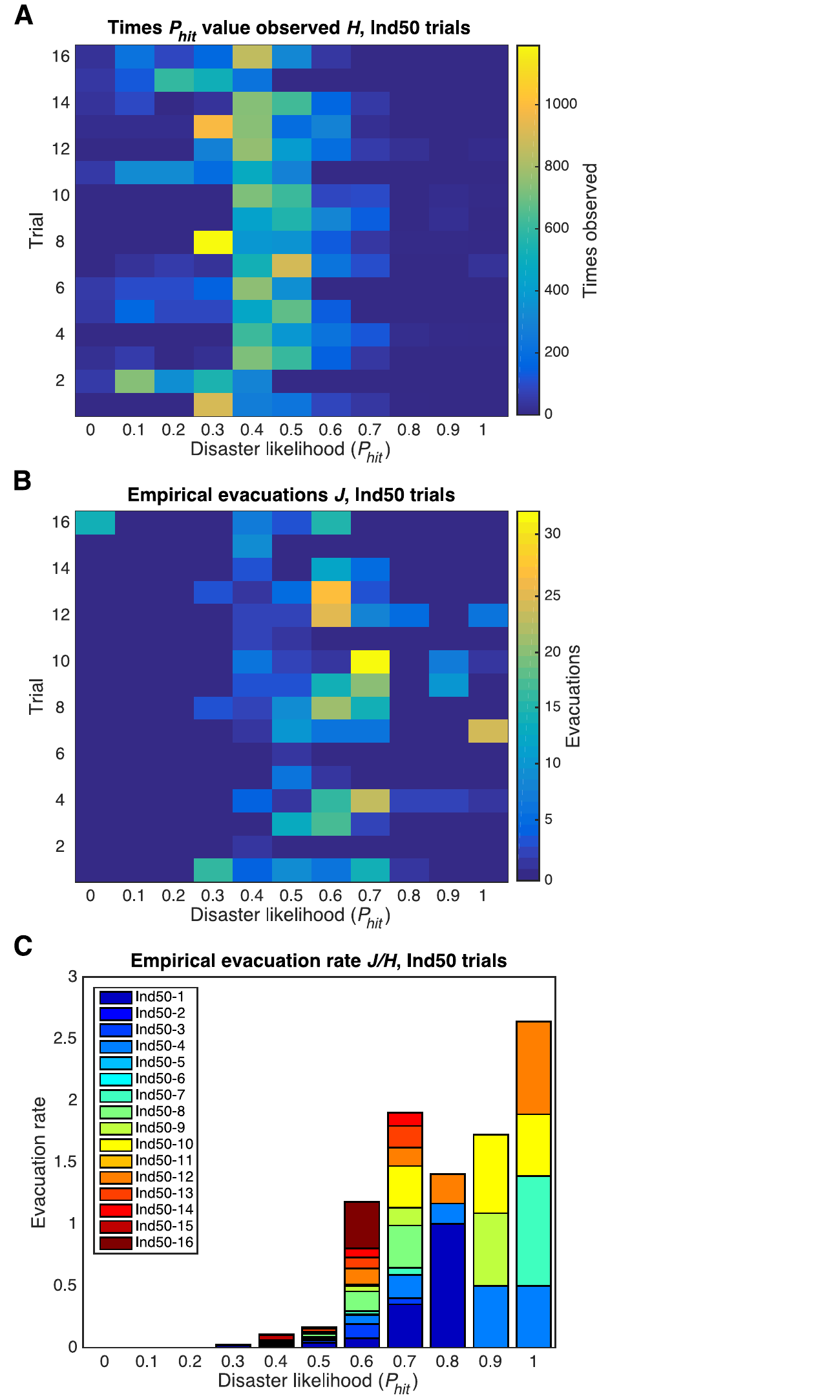}
\caption{{\bf Evacuation rates as a function of $\phit$.} A: Depiction of the number of times each $\phit$ value $\nu$ was observed, denoted $H_{\nu}$. B: Depiction of the number of evacuations $J_{\nu}$ as a function of $\phit$. C: Plot of the observed rate of evacuations $J_{\nu}/H_{\nu}$ for each of the 16 individual trials with 50 initial shelter spaces.}
\label{fig:empiricalrate}
\end{figure}

The values of $H_{\nu}$ and $J_{\nu}$ as a function of $\phit$ are plotted in Figs.~\ref{fig:empiricalrate}A and~\ref{fig:empiricalrate}B, as well as the empirically observed evacuation rate, $J_{\nu}/H_{\nu}$, in Fig.~\ref{fig:empiricalrate}C. $J_{\nu}/H_{\nu}$ corresponds to $(\alpha - 1)/(\alpha + \beta - 2)$, the mode of the Beta distribution with parameters $\alpha$ and $\beta$.

\begin{figure}[h!]
\includegraphics{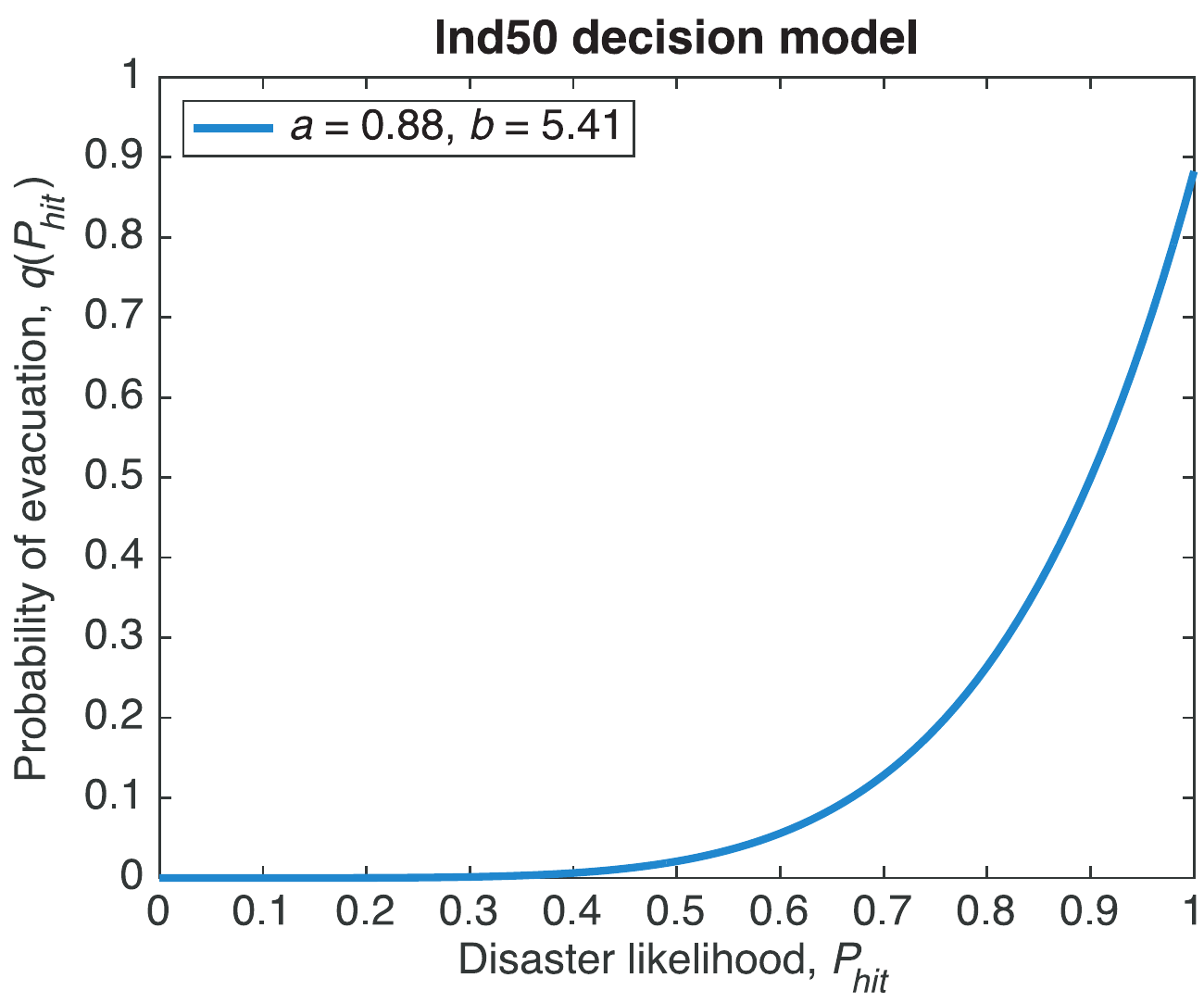}
\caption{{\bf Individual decision model.} A plot of the decision model $q$, which represents the probability that an individual will decide to evacuate, as a function of the disaster likelihood $\phit$, fitted to the observed behavior plotted in Fig.~\ref{fig:empiricalrate}. Here, $q$ has a power-law form described in Eq.~\eqref{eq:decmodel}, with parameters $a = 0.88$ and $b = 5.41$. This function forms the basis for the evacuation predictions shown in Figs.~\ref{fig:onetrial} and~\ref{fig:indivtrials}, isolating the influence of the disaster likelihood on evacuation decisions without incorporating information about the precise time steps on which decisions occur.}
\label{fig:powermodel}
\end{figure}

The most commonly-observed $\phit$ value is 0.4, since the majority of trials start at $\phit = 0.4$, whereas the most common $\phit$ at evacuation is 0.6. The evacuation rate does not increase monotonically with $\phit$; rather, there are two local maxima, located at $\phit = 0.7$ and $\phit = 1$. The evacuation rate is zero or nearly zero for $\phit \in \{0, 0.1, 0.2, 0.3, 0.4\}$. 
Minimizing expression~\eqref{eq:ll} with respect to the parameters of Eq.~\eqref{eq:decmodel}, we obtain the fitted values $a = 0.88 \pm 0.03$ and $b = 5.41 \pm 0.27$. As described in the previous section, $b$ represents the concavity of the function, while the value of $a$ indicates that probability of evacuation approaches a maximum of $0.88$ as the disaster likelihood reaches $1$ (a definite disaster strike). This fitted decision ``strategy,'' denoted $q$, gives the probability of evacuation for each participant as a function of $\phit$ and is plotted in Fig.~\ref{fig:powermodel}. This individual strategy forms the basis of the model, isolating the influence of the broadcast value of $\phit$ on evacuation decisions.

Using this basic best-fit decision strategy, we then solve the master equation given in Eq.~\ref{eq:mastereq}, using the $\phit(t)$ trajectory of each trial to obtain the full probability distribution of evacuations at each time step of that trial. We determine the model predictions by calculating the mean number of cumulative evacuations as a function of time. Comparing the predicted evacuations with the empirical observations, we obtain a root-mean-squared-error (RMSE) of 6.8 evacuations per time step on individual trials with 50 initial shelter spaces, averaged over time steps and trials. 

\begin{figure}[h!]
\includegraphics{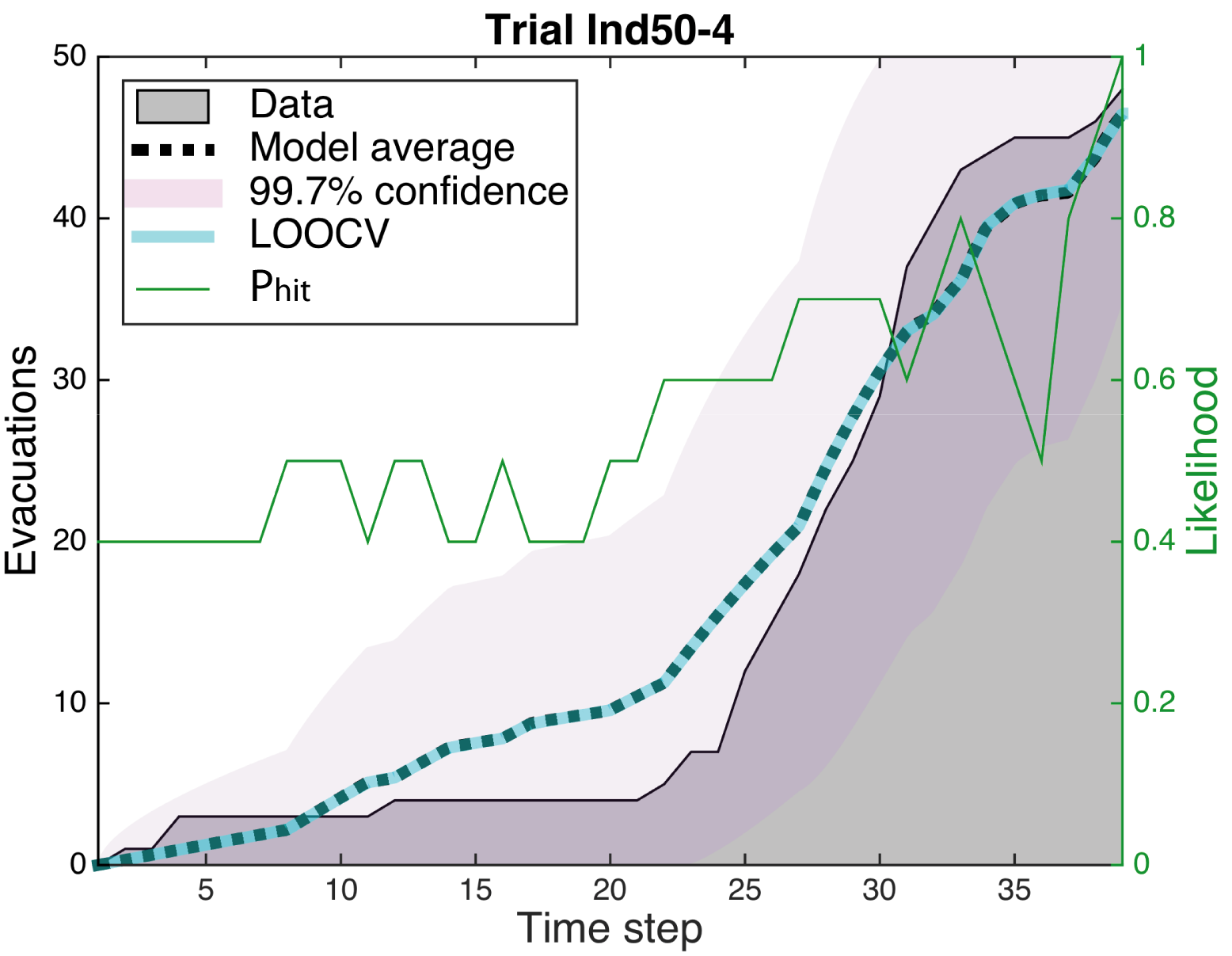}
\caption{{\bf Model prediction of population evacuation behavior.} Comparison between empirical behavior and model predictions for the 4th individual trial with 50 initial shelter spaces. The cumulative number of evacuations observed over time during this trial is plotted as a gray shaded area. The mean number of predicted evacuations (dashed black line) and the 99.7\% confidence interval (magenta area) are also shown, calculated from the full probability distribution of evacuations generated by a model fit to all individual trials with 50 shelter spaces. Also shown are the LOOCV prediction (blue line), showing the mean evacuations predicted by a model fit to all such trials except the one shown here, and the $\phit$ trajectory (green line), or the likelihood of the disaster striking as seen by participants during the trial.}
\label{fig:onetrial}
\end{figure}

An example of the model prediction is shown in Fig.~\ref{fig:onetrial} for the 4th trial of this set. In this figure, the gray shaded area shows the cumulative evacuations observed during the experiment, as a function of time, while the dotted line shows the expected behavior -- i.e., the mean of the distribution of cumulative evacuations at each time step -- as predicted by the model fit to all such individual trials. The shaded magenta area gives the 99.7\% ($3\sigma$) confidence interval for this model prediction, also computed from the full probability distribution of evacuations given by the model. The light blue line depicts the expected model behavior as predicted by a model fit to all individual trials except the trial shown; its close similarity to the original model fit to all trials (dotted line) shows that the performance of the model on this trial is not strongly affected by overfitting, and that the performance can be expected to generalize, as discussed further below. Finally, the green line shows the $\phit$ value seen by participants on each time step of the trial. In this particular trial, the model predicts a slightly larger rate of evacuation than is observed for approximately the first 30 time steps of the trial, but closely replicates the sharp increase in evacuations following the jump in $\phit$ from $0.4$ to $0.7$ that occurs around time step 20. A leveling-off of evacuations occurs around time step $33$ in both the data and model prediction, when $\phit$ dips from $0.8$ to $0.5$, followed by a small uptick in evacuations as $\phit$ increases again to $1$, after which the trial ends.

To determine if the model is overfitting the data, we performed leave-one-out cross-validation (LOOCV). We found that the LOOCV predictions were very close to that of the original model trained on all 16 trials. Specifically, the RMSE of the LOOCV predictions, which measures the disparity between predicted and observed cumulative evacuations over all time steps, is 7.1 evacuations on average over all 16 trials. This demonstrates that the model is expected to generalize to unseen data, making similarly accurate predictions on new individual trials, assuming they are run according to the same parameters. 

\begin{figure*}[ht]
\includegraphics[width=\linewidth]{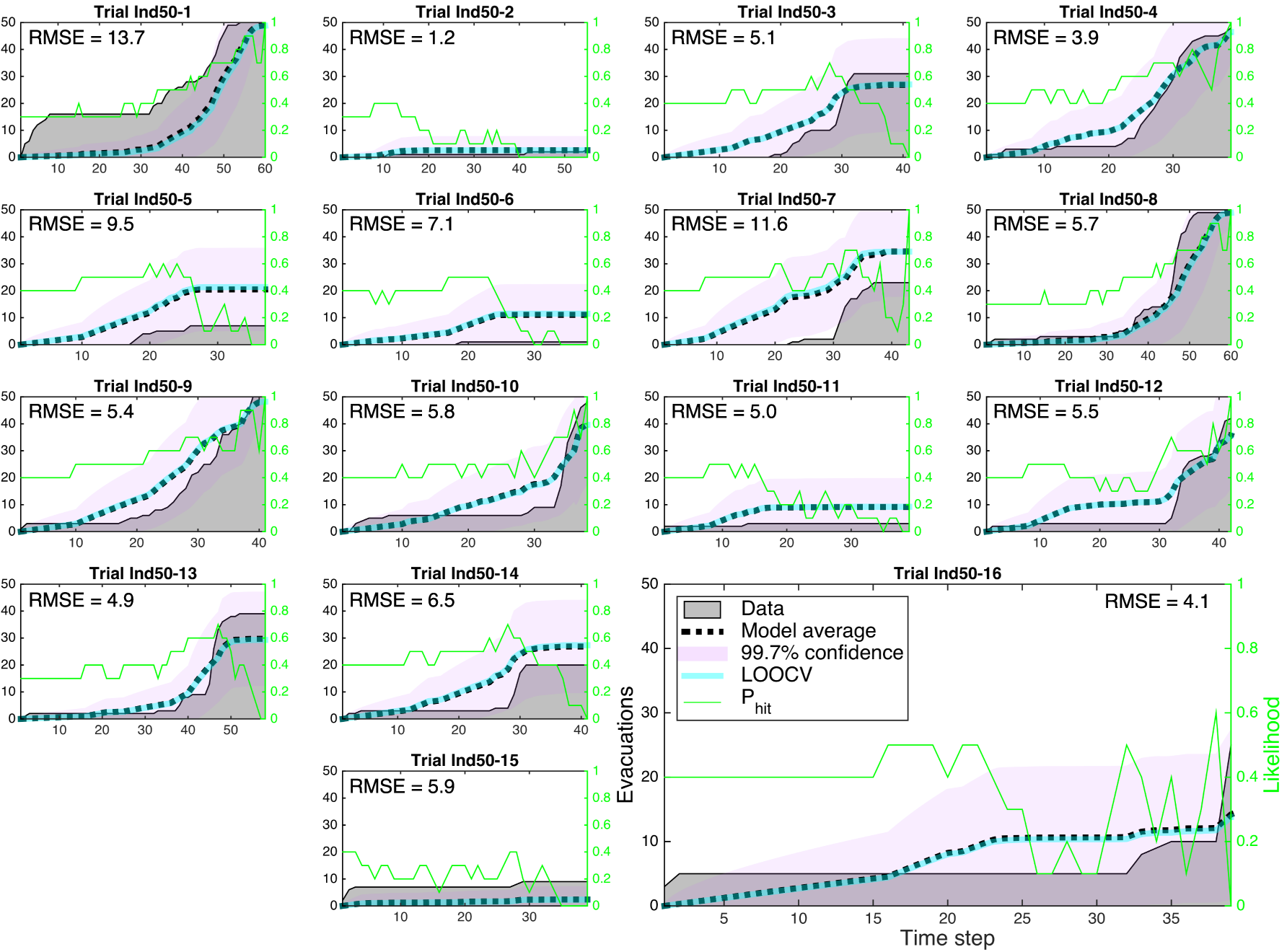}
\caption{{\bf Model predictions of population behavior on individual decision trials.} The predicted cumulative number of evacuations as a function of time (black dashed lines) and 99.7\% confidence intervals (magenta area) for each Ind50 trial are plotted alongside empirical values (gray area). $\phit$ trajectories are shown in green. The leave-one-out cross-validation (LOOCV) predictions are also plotted in blue.}
\label{fig:indivtrials}
\end{figure*}

The original model predictions, the 99.7\% confidence intervals, and the LOOCV predictions are plotted for each of the 16 trials in Fig.~\ref{fig:indivtrials}, compared to the empirical evacuation rates measured in the experiment. Note that these model predictions do not use the time step as an explicit factor in prediction, but are based only upon the influence of instantaneous reports of the value of $\phit$ seen by the participants, regardless of whether they were seen early or late in a trial.

\begin{figure}[ht]
\includegraphics{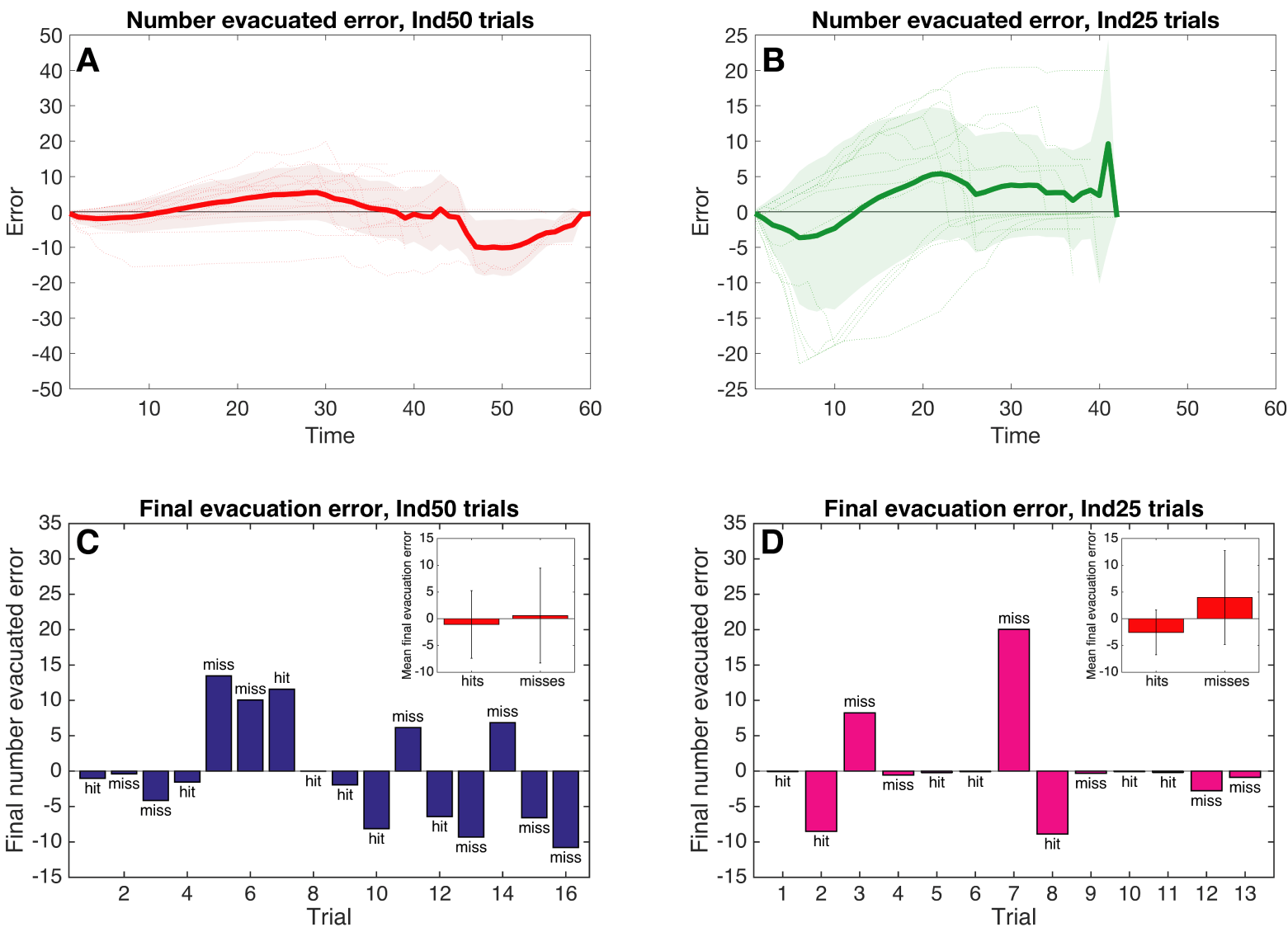}
\caption{{\bf Model prediction error for individual decision trials.} A: The error in number evacuated as a function of time for each of the 16 individual trials with 50 initial shelter spaces (dashed red lines) and the errors averaged over all trials still active at each time step (thick red line). B: The error in number evacuated as a function of time for each of the 13 individual trials with 25 initial shelter spaces (dashed green lines) and the errors averaged over all trials still active at each time step (thick green line). C: The final error in number evacuated at the end of each trial for the 16 individual trials with 50 initial shelter spaces; the averaged final evacuation error (inset) for disaster hits and disaster misses. D: The final error in number evacuated at the end of each trial for the 13 individual trials with 25 initial shelter spaces; the averaged final evacuation error (inset) for disaster hits and disaster misses.}
\label{fig:indverrors}
\end{figure}

Fig.~\ref{fig:indverrors}A plots the error in number evacuated as a function of time for each trial. On average, the model underestimates the number of evacuations in the early stages of the trials, between time steps 0 and 10; overestimates the number of evacuations between time steps 10 and 35; and ultimately underestimates evacuations for the remaining time steps.  Fig.~\ref{fig:indverrors}C plots the final evacuation errors, i.e. the difference between the predicted and observed number of evacuees at the end of each trial, with the inset displaying the errors averaged over trials in which the disaster strikes and trials in which the disaster does not strike. On average, the model slightly underpredicts the final number of evacuations in trials where the disaster hits, and very slightly overpredicts the evacuations in trials where the disaster misses. The variance is larger in disaster misses than in disaster hits.

\begin{figure}[h!]
\includegraphics[width=\linewidth]{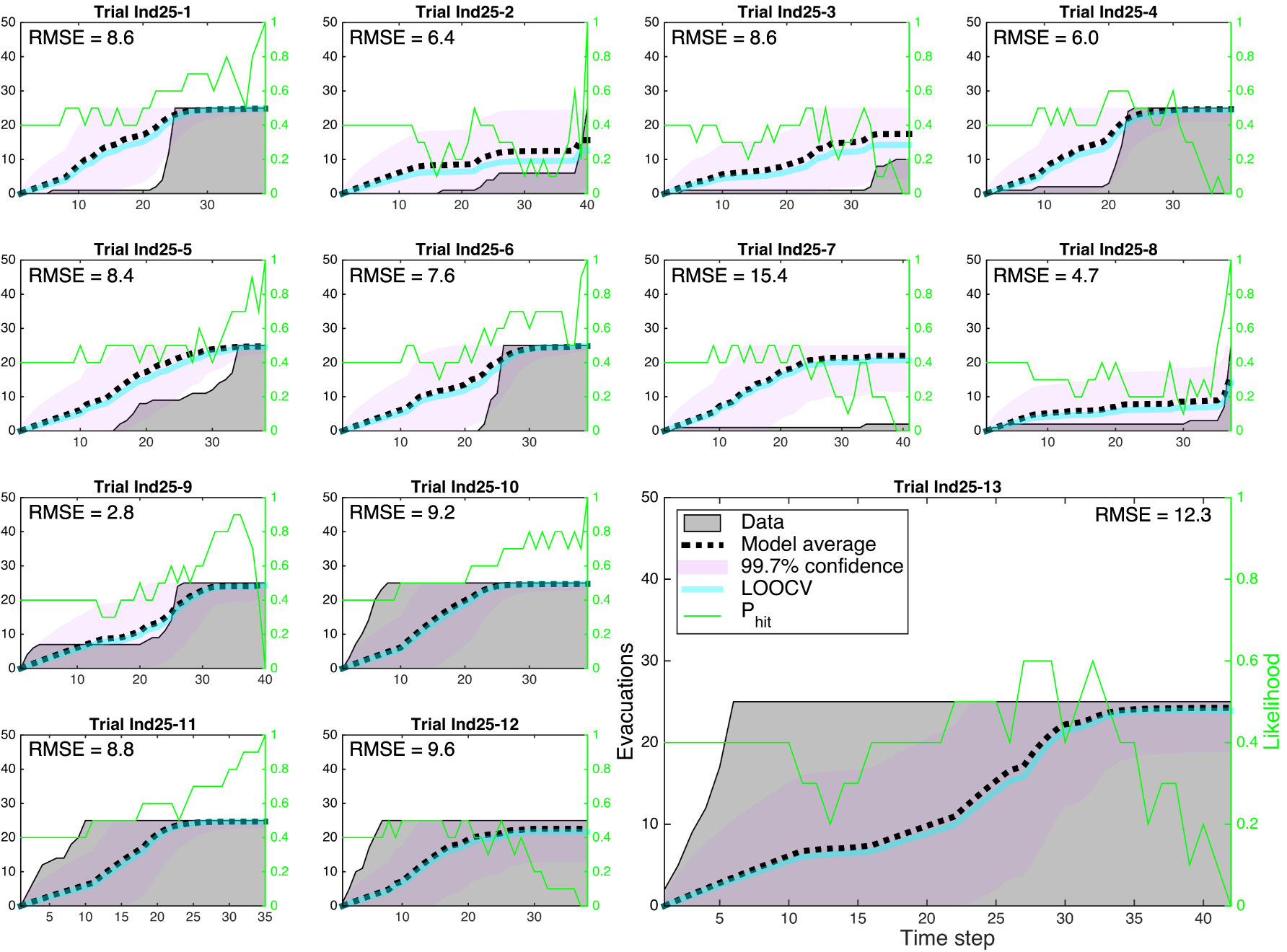}
\caption{{\bf Model predictions of population behavior on individual decision trials with limited shelter capacity.} For each individual decision trial with an initial shelter space of 25, the cumulative number of evacuations predicted by the model is plotted as a function of time (magenta dashed lines), along with 99.7\% confidence intervals (magenta area). In comparison, empirical evacuations measured in the experiment are shown as a shaded gray area, and LOOCV model predictions are plotted as blue lines. $\phit$ trajectories are shown in green.}
\label{fig:shelter25}
\end{figure}

\textbf{Influence of initial shelter capacity.}
We extend the model to include a dependence on the initial shelter capacity, $s$, in addition to $\phit$, using Eq.~\eqref{eq:sp} from the Methods section. The parameter $a$ is constrained to have the same value as the model trained on Ind50 trials, $a = 0.88$. We fit this model to the thirteen Ind25 trials and obtain the remaining parameter of Eq.~\eqref{eq:sp}, $c 
= 5.55 \pm 2.64$.
The RMSE between prediction and data is 8.88 averaged over all time steps and all trials, and the RMSE for the LOOCV predictions is 7.56. The model predictions, confidence intervals, data, LOOCV predictions, and RMSE values (averaged over all time steps for each trial) are plotted for these trials in Fig.~\ref{fig:shelter25}. The model predictions generally do not fit the shape of the empirical evacuation trajectories as well as the individual decision model for Ind50 trials, failing to predict the sharp increases in evacuations that often occur following rapid increases in $\phit$. Fig.~\ref{fig:indverrors}B shows the number evacuated error as a function of time for each trial. Like the previous $\phit$-only model (Fig.~\ref{fig:indverrors}A), the shelter-space-dependent model underpredicts evacuations at early time steps and overpredicts evacuations at moderate time steps, but is on average accurately predictive of evacuations at late time steps. Fig.~\ref{fig:indverrors}D shows the final number evacuated error for each trial, as well as the errors averaged over disaster hits and misses. Like the $\phit$-only model (Fig.~\ref{fig:indverrors}C), the model underpredicts the average final number evacuated for disaster hits and (more significantly) overpredicts the average final evacuations for disaster misses, with a larger variance for misses.

For the fifteen Ind5 trials, competition for shelter space appears to be more influential in decision-making than $\phit$. All evacuations occurred by time step $t=3$, even though the trial is guaranteed to continue until at least $t=20$. In contrast, in Ind25 trials, the shelter often did not fill up by $t=20$ (Trials 1, 2, 4, 5, 6, 8, 9) or did not fill up at all (Trials 3 and 7, where $\phit$ remained less than 0.6). (However, the shelter filled up by $t=10$ for the final four Ind25 trials, despite $\phit$ remaining at 0.5 or lower.) 
Furthermore, all evacuations in Ind5 trials occurred at $\phit = 0.3$ or $0.4$; therefore, there is not enough data to determine the effect of $\phit$ on behavior.
Participants likely evacuate early in Ind5 trials to avoid being locked out of the shelter. Being in shelter during a disaster miss gives a smaller loss than being at home during a disaster strike; participants would have lost 4 points for remaining home during a strike versus 2 points for evacuating during a miss. Participants may weigh this against the very high level of competition for shelter space and decide that preemptively evacuating to ensure a space is the best strategy, especially after several trials in which other participants are clearly adopting this strategy.
Since the behavior during trials with 5  shelter spaces changes very little as a function of $\phit$, group size, or group decision protocol, the remainder of this work will focus on trials with 50 or 25 initial shelter spaces.
\textbf{Comparison of individual and group behavior.}
Having characterized the accuracy of the Markov model in predicting population-level behavior when each member of the population is making an individual decision, we now extend this model to explore how decision dynamics differ when individuals must make collaborative evacuation decisions in groups of various sizes. We also investigate the influence of different decision protocols on the dynamics of these groups. We initially focus on situations with shelter space available for all participants, and follow this with an investigation of the effect of limited shelter space upon the results.

\begin{figure*}[ht]
\includegraphics[width=\linewidth]{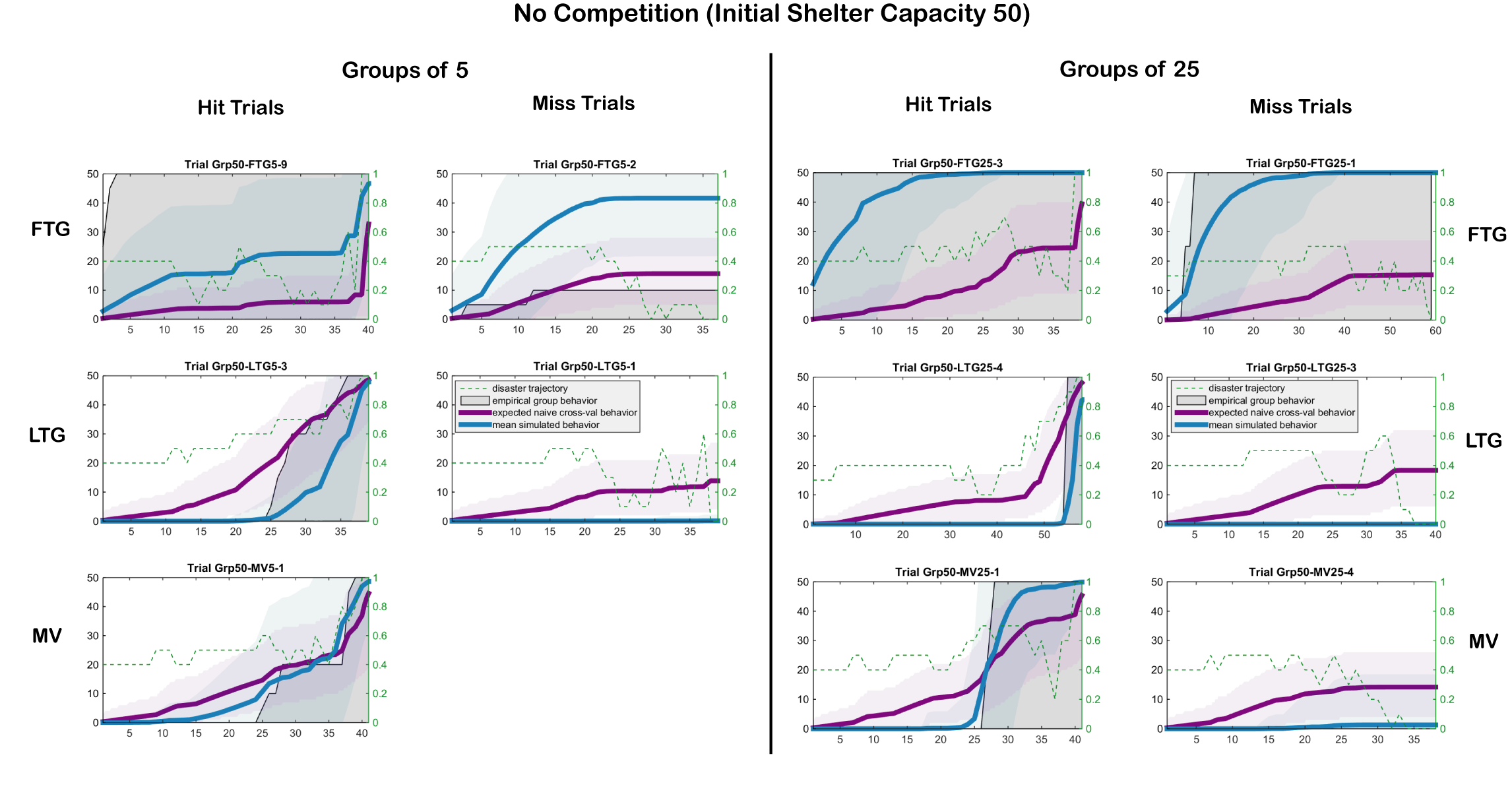}
\caption{\textbf{Individual-group differences in example trials.} Empirical cumulative evacuations for individual trials are shown as gray shaded areas, compared to na\"ive cross-validation model predictions (purple lines, with 99.7\% confidence intervals in magenta) and grouped-individual simulations (blue lines, with 99.7\% confidence intervals in light blue). Green dashed lines show likelihood of disaster hitting. A representative example trial is displayed for each scenario of a given group size, decision protocol, and eventual trial outcome. All trials shown have initial shelter space of 50.}
\label{fig:ind_grp_trials50}
\end{figure*}

We first quantify the extent to which the population-level behavior in individual scenarios differs from the behavior in a group decision scenario, using the ``na\"ive cross-validation'' method of comparing group and individual behavior. To do this, we first fit the stochastic decision model for independent individual decisions to individual trials, as in Fig.~\ref{fig:powermodel}; we then test the accuracy of these individual fit parameters in predicting group behavior in scenarios with the same shelter capacity. We do not expect particularly good prediction from this model, since the na\"{i}ve cross-validation does not explicitly account for the enforced group structure of the scenarios, but simply compares the total number of empirical group-scenario evacuations to the expected number of evacuations under the individual model. However, the procedure gives a baseline quantification of the differences between group and individual population behavior. Na\"{i}ve cross-validation predictions are shown as purple curves in example trials in Fig.~\ref{fig:ind_grp_trials50}; magenta shading indicates the 99.7\% confidence interval computed from the predicted evacuation probability distribution.

Na\"{i}ve cross-validation reveals clear differences between group and individual behavior at a population level. We next perform grouped-individual simulations to distinguish whether these differences stem merely from the group constraint imposed upon the evacuation dynamics, or whether individuals make changes to their personal decision strategies when they know they are in a group decision scenario. These simulations use the Gillespie algorithm to generate sample instances of the population evacuation behavior, assuming that each participant evacuates according to the best-fit decision strategy for individual trials, but enforcing the group decision constraints in group-5 and group-25 scenarios. If participants assigned to a group decision framework use the same decision strategies they would have used in an individual scenario, we would expect these simulations to give statistically identical results to the empirically observed behavior. Grouped-individual simulation results, averaged over 1000 simulated trial instances, are depicted as blue curves in Fig.~\ref{fig:ind_grp_trials50}; light blue shading indicates the 3$\sigma$ (99.7\%) confidence interval from the distribution of simulation instances.

\begin{figure}[ht]
\includegraphics{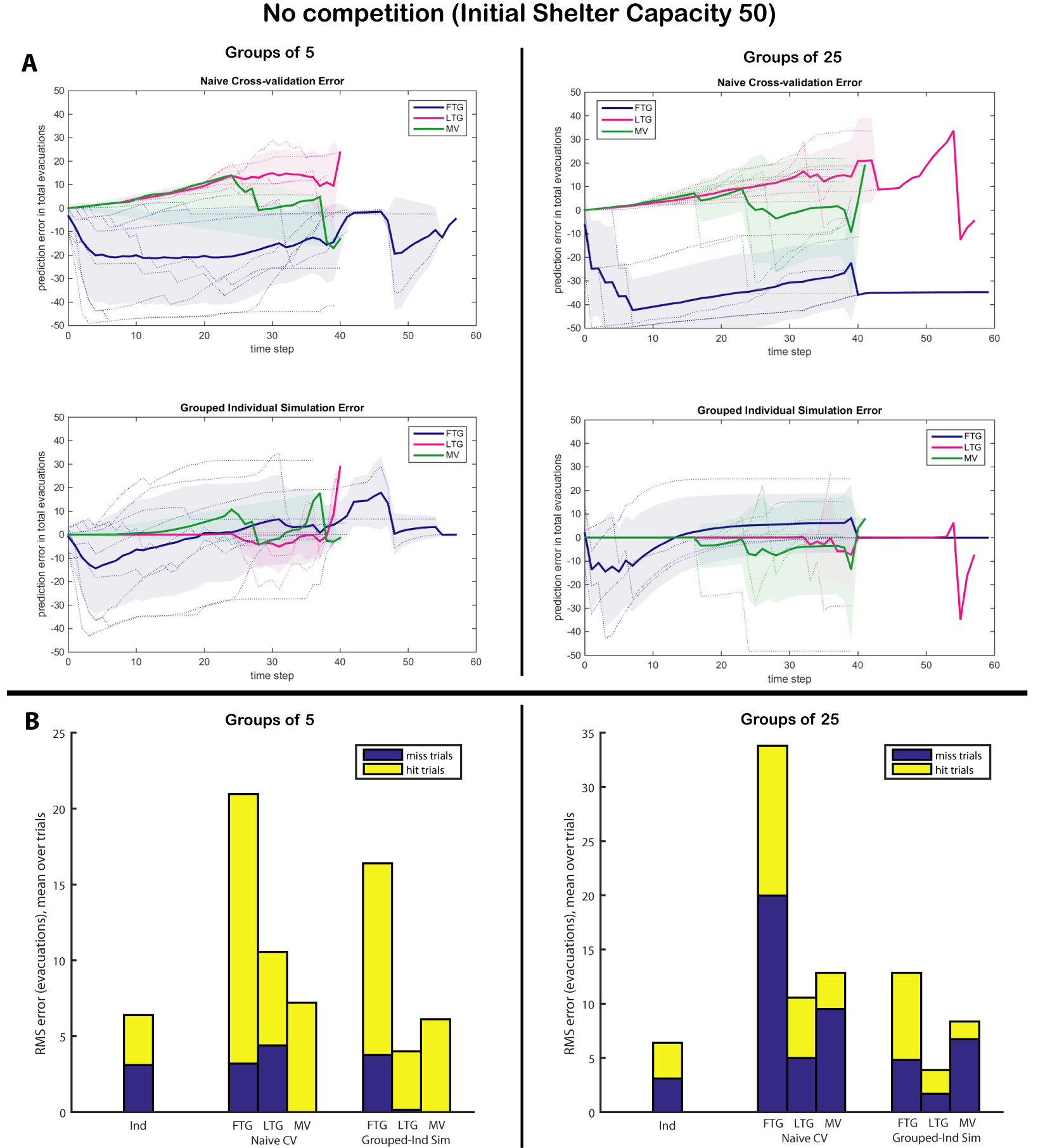}
\caption{\textbf{Individual-group differences and prediction errors.} A: Prediction error on each time step for na\"ive cross-validation (top panels) and grouped individual simulations (bottom panels). All group trials with 50 initial shelter spaces are shown, separated into FTG (blue), MV (green), and LTG (pink) decision protocols. Each dotted line represents a single trial, and solid lines represent an average over trials. B: Total root-mean-square error (RMSE) by decision protocol for na\"ive cross-validation and grouped individual simulations, on group-5 (left) and group-25 (right) trials. Each bar shows RMSE averaged over all trials in the given category, and colors represent the contribution of error from hit versus miss trials, weighted by both number of each type of trial and the size of the errors contributed. All trials included have initial shelter space of 50.}
\label{fig:ind_grp_errors50}
\end{figure}

\begin{figure*}[ht]
\includegraphics[width=\linewidth]{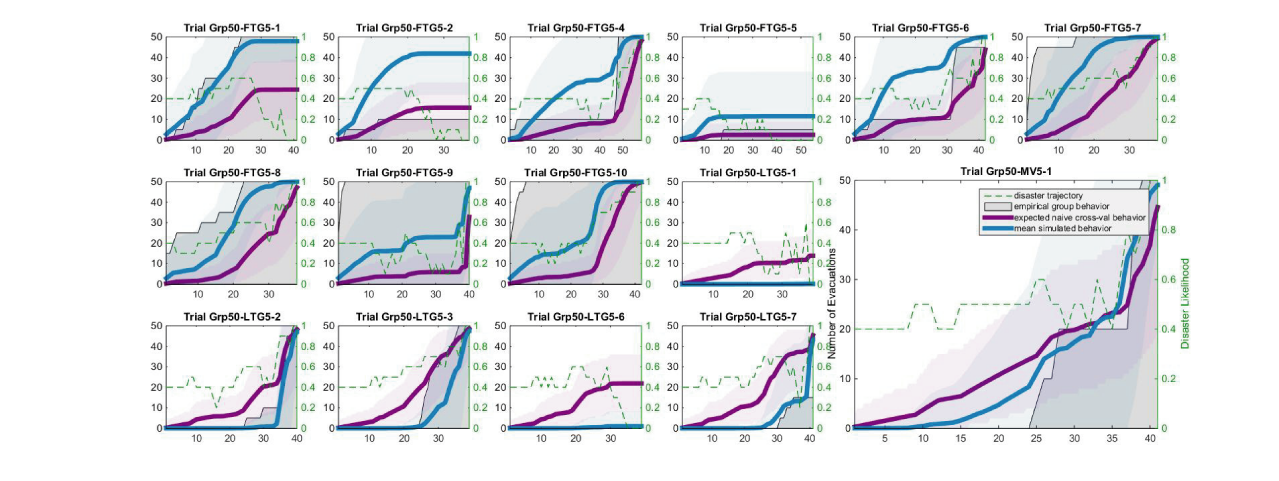}
\caption{\textbf{Individual versus group decision strategies in group-5 trials.} Empirically observed cumulative evacuations (gray shaded area) on group-5 trials with initial shelter space of 50, compared to two different models of group behavior. Na\"ive cross-validation predictions, or the expected value of behavior predicted solely by the best fit individual decision strategy with no group structure, are shown in solid purple, with 99.7\% confidence interval in magenta. Grouped-individual simulation results, representing the behavior of groups assuming all group members follow the individual best-fit strategy, are shown in solid blue (average over 1000 simulations), with confidence interval in light blue. All three decision protocols -- first-to-go (FTG), majority-vote (MV), and last-to-go (LTG) -- are shown. Selected example trials from all three protocols are highlighted in Figure \ref{fig:ind_grp_trials50} in the main manuscript.}
\label{fig:alltrials_50_5}
\end{figure*}

\begin{figure*}[ht]
\includegraphics[width=\linewidth]{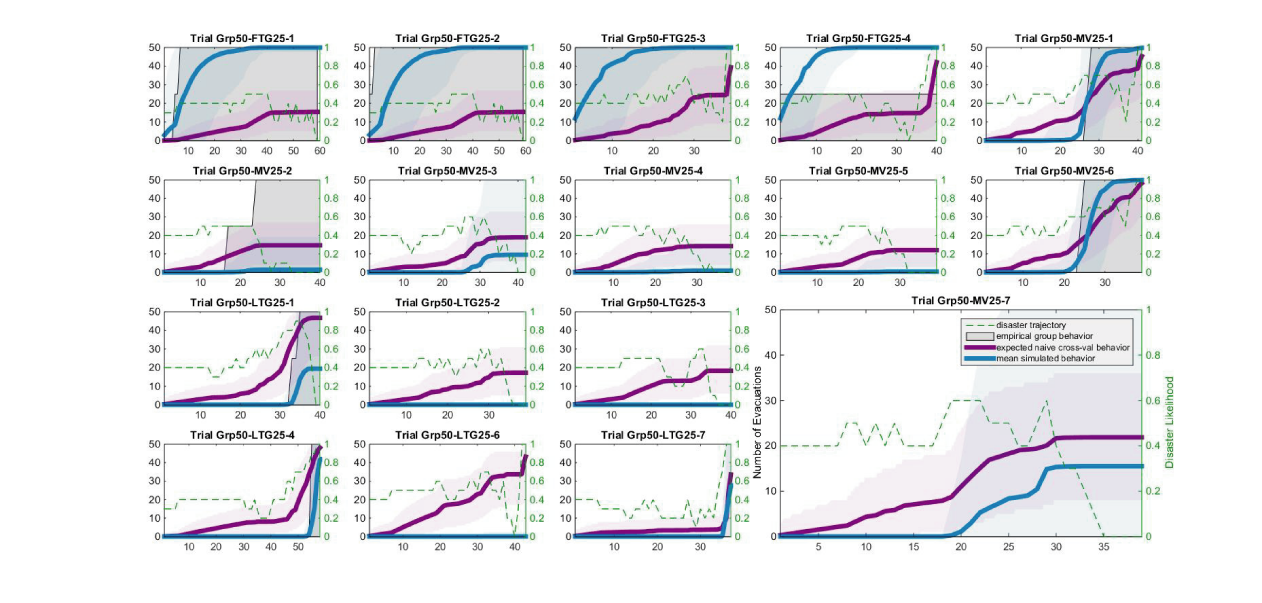}
\caption{\textbf{Individual versus group decision strategies in group-25 trials.} Empirically observed cumulative evacuations (gray shaded area) on group-25 trials with initial shelter space of 50, again compared to na\"ive cross-validation predictions (shown in solid purple, with 99.7\% confidence interval in magenta) and grouped-individual simulation results (average over 1000 simulations shown in solid blue, with confidence interval in light blue). All three decision protocols -- first-to-go (FTG), majority-vote (MV), and last-to-go (LTG) -- are shown. Selected example trials from all three protocols are highlighted in Figure \ref{fig:ind_grp_trials50} in the main manuscript.}
\label{fig:alltrials_50_25}
\end{figure*}

Fig.~\ref{fig:ind_grp_trials50} compares empirically observed evacuation behavior (shaded gray area) to both na\"{i}ve cross-validation (solid purple line) and grouped-individual simulations (solid blue line) across different group sizes and decision protocols. Two representative example trials are shown for each type trial category, one in which the disaster ultimately hits the population and one in which it misses. Note that variation between trials does exist; similar figures for all trials are included in Figs.~\ref{fig:alltrials_50_5} and~\ref{fig:alltrials_50_25}. Fig.~\ref{fig:ind_grp_errors50}A shows the errors in evacuation prediction for na\"{i}ve cross-validation (top panel) and grouped-individual simulation (bottom panel), including all trials (dotted lines) as well as the average over all trials of each decision protocol (solid lines).

Evacuation behavior at a population level shows some clear changes between individual to group-5 scenarios. Most strikingly, the models severely underestimate early evacuations in the depicted FTG hit trial; this error is most severe with the na\"{i}ve cross-validation model, which approximates what a population of individuals would have done, but the grouped-individual simulations do not significantly improve on this in group-5 trials, despite taking group structure into account. This indicates that participants have fundamentally changed their strategies for group trials, responding to factors not included in these models. On both MV and LTG hit trials, the model estimate is much closer to observed behavior, indicating that changes in strategy are less extreme and the participants hew closer to their individual strategy on these trials. However, in miss trials, this trend is somewhat reversed: actual evacuation numbers are lower, and the models tended to overestimate evacuation. Interestingly, the group-5 FTG trials from Fig.~\ref{fig:ind_grp_trials50} have very similar $\phit$ trajectories regardless of their eventual hit/miss outcome; the drastic difference in evacuations between the two trials is likely due to considerations beyond $\phit$, group decision protocol, and size, for example the outcomes of the immediately preceding trials.

In group-25 trials, underestimation of early evacuation in FTG trials by na\"{i}ve cross-validation grows even worse, as do overall na\"{i}ve predictions on MV and LTG trials. However, the grouped-individual simulations are better able to predict the evacuation outcomes in most scenarios. This could indicate that participants tend to stay closer to their individual strategies in large groups more than they do in small groups, or it could reflect the tendency of large groups with a rigid decision protocol to amplify extreme behavior such as early evacuations, making it easier for the model to predict the entire group evacuation time based only on predicting the behavior of a few extreme individuals.

Fig.~\ref{fig:ind_grp_errors50}B plots the total root mean squared error of each scenario, averaged over time steps and trials. Each bar is divided proportionally to depict the amount of error contributed by hit versus miss trials (weighted by number of trials of each type as well as their error contribution). The generally large errors in na\"{i}ve cross-validation reflect a baseline measurement of the differences between group and individual dynamics on the same trials. When grouped, populations tend to show much more altered overall behavior with FTG protocols than they do under LTG and MV protocols. The grouped-individual errors reflect changes in individual strategy, not caused by the group structure of the decision.

\begin{figure*}[ht]
\includegraphics[width=\linewidth]{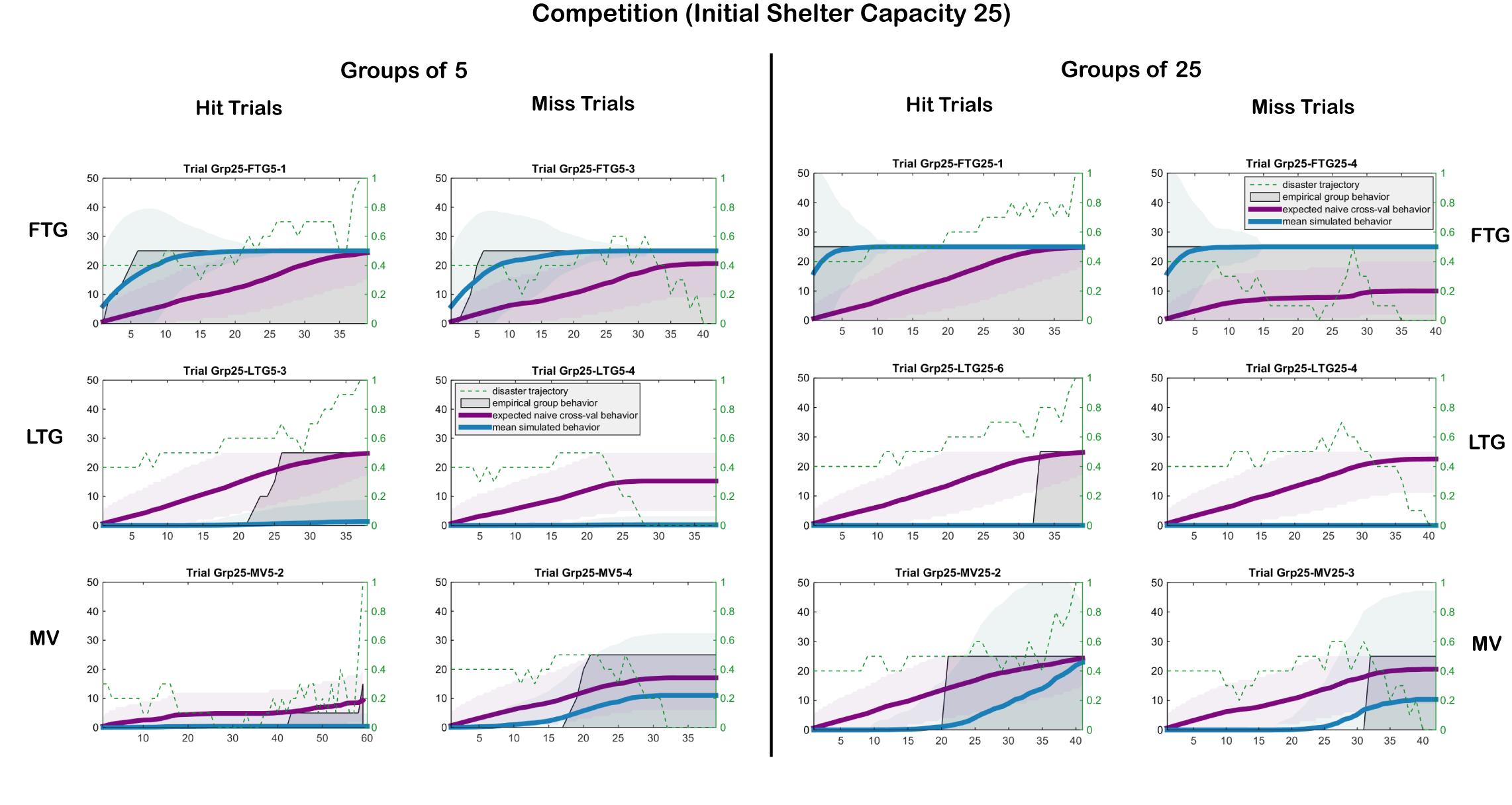}
\caption{\textbf{Individual-group differences in example trials under competition.} Empirical cumulative evacuations for individual trials are shown as gray shaded areas, compared to na\"ive cross-validation predictions (purple lines, with 99.7\% confidence intervals in magenta) and grouped-individual simulations (average over 1000 simulations shown as blue lines, with 99.7\% confidence intervals shown in light blue). Green dashed lines show likelihood of disaster hitting. A representative example trial is displayed for each scenario of a given group size, decision protocol, and eventual trial outcome. All trials shown have initial shelter space of 25.}
\label{fig:ind_grp_trials25}
\end{figure*}

\begin{figure}[ht]
\includegraphics{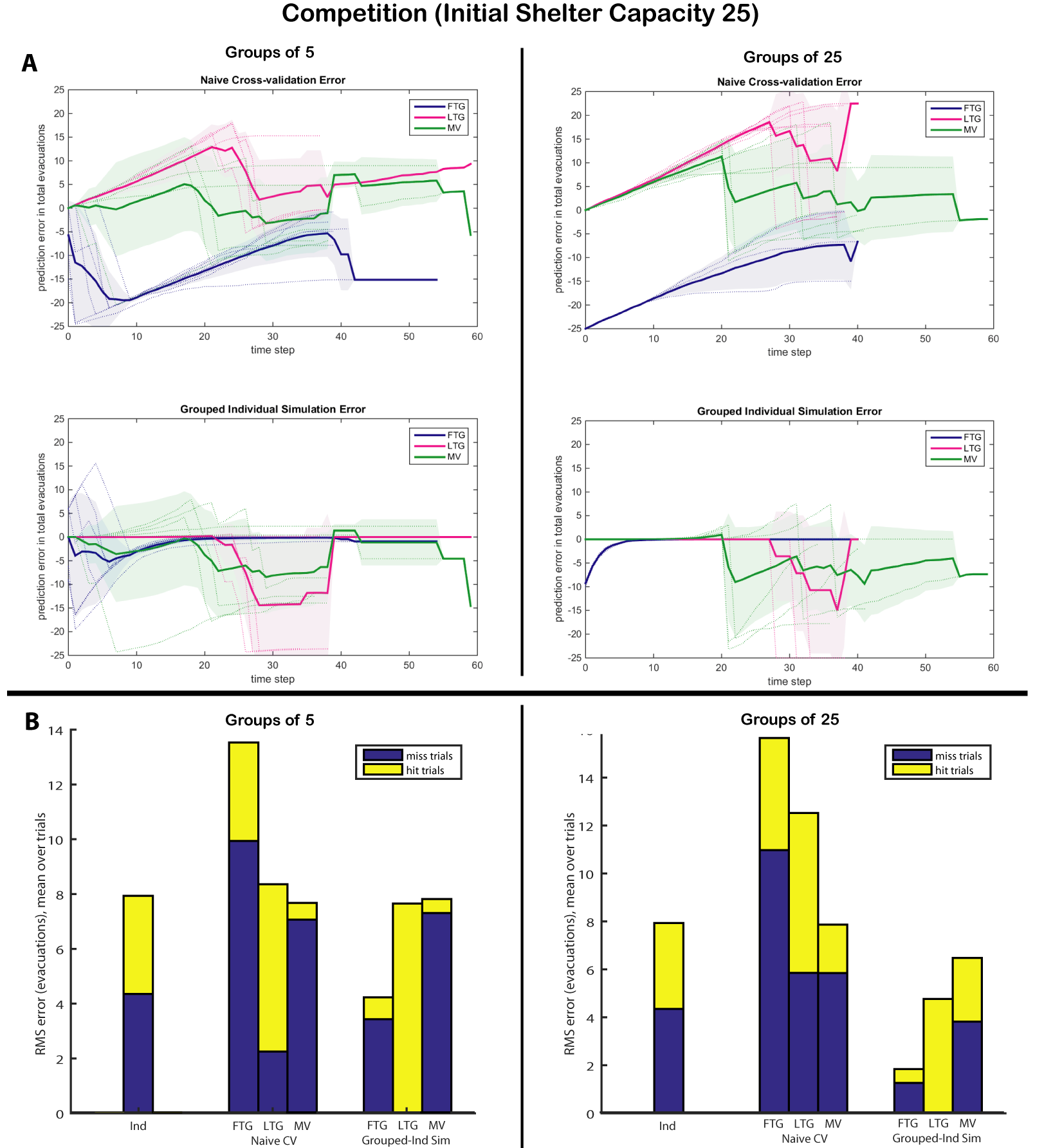}
\caption{\textbf{Individual-group differences and prediction errors under competition.} A: Prediction error on each time step for na\"ive cross-validation (top panels) and grouped individual simulations (bottom panels). All group trials with 25 initial shelter spaces are shown, separated into FTG (blue), MV (green), and LTG (pink) decision protocols. Each dotted line represents a single trial, and solid lines represent an average over trials, including only those trials still active at each given time step. B: Root-mean-squared error (RMSE) by decision protocol for na\"ive cross-validation and grouped individual simulations, on group-5 (left) and group-25 (right) trials. Each bar shows RMSE averaged over all trials in the given category, and colors represent the contribution of error from hit versus miss trials, weighted by both number of each type of trial and the size of the error. All trials included have initial shelter space of 25.}
\label{fig:ind_grp_errors25}
\end{figure}

Finally, we examine the differences between group and individual behavior in the case of limited shelter space. Figs.~\ref{fig:ind_grp_trials25} and~\ref{fig:ind_grp_errors25} show group predictions from na\"ive cross-validation and grouped-individual simulations, compared with empirical evacuation behavior, in scenarios with only 25 initial shelter spaces available.

As we might expect in situations where there is competition for shelter space, FTG protocols see the shelter space filled almost immediately in both group-5 and group-25 trials, almost regardless of the $\phit$ trajectories. The na\"{i}ve model predicts that individuals would take longer to fill the shelter without group structure, but the grouped-individual simulations predict the early evacuation quite well, suggesting that individual strategies remain similar in FTG group trials, and the resulting evacuation behavior is what would be expected from simply grouping participants who are using strategies from individual trials. 

Interestingly, while grouped-individual simulations accurately predict the near-immediate filling of the shelter in FTG trials, they often fail to predict the filling of the shelter in LTG trials, and underestimate it on MV trials as well. In several instances, even under a LTG protocol, multiple groups in the actual experiment unanimously agree to evacuate as the disaster likelihood climbs, even when individual decisions alone would not predict unanimous group evacuation. In these cases a change to individual strategy, perhaps caused by awareness of being in a group or even a cascading effect of many successive decisions, allows the entire group to come to an evacuation consensus in practice, which serves them well in avoiding an eventual disaster.
\textbf{Optimal evacuation strategies.}
Solving for the \textit{static optimal} strategy gives us an interesting window into the trial setup. With the 2-parameter power-law model, we look for the set of parameters that produces an expected optimal static score closest to the best possible score in practice, $342$ (over 48 unique $\phit$ trajectories). Note that this ``best possible'' score is composed of perfect evacuation decisions on all trials: when the disaster hits, this perfect strategy is able to always evacuate \textit{with certainty} -- i.e., the probability of evacuation is unity -- earning a score of $4$ for property loss; when the disaster misses, the perfect strategy always stays home \textit{with certainty}, earning a score of $10$ (no losses).

\begin{figure}[ht]
\includegraphics{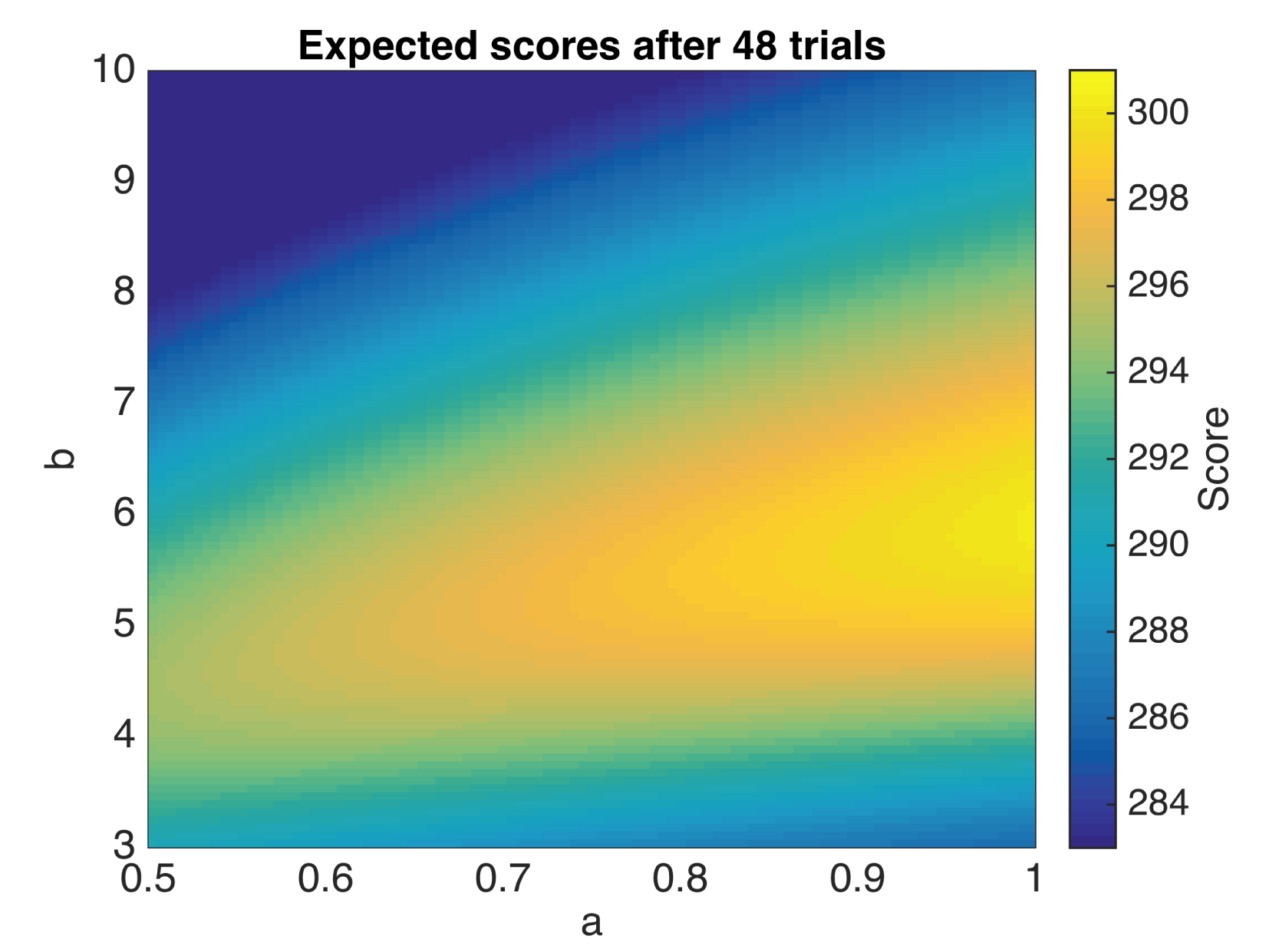}
\caption{{\bf Static optimal strategy.} Color denotes the expected score over all $48$ unique $\phit$ trajectories, as a function of the decision model parameters $a$ and $b$. We use a decision model of the form $q(\phit) = a \left(\phit\right)^b$ (see Eq.~\eqref{eq:decmodel}), and assume that the strategy is the same for all $48$ trials. The optimal score occurs for $a = 1$, $b = 5.90$, with near-optimal scores occurring in the range $0.75 \leq a \leq 1$, $4.5 \leq b \leq 7$.}
\label{fig:param_space_opt_ind50}
\end{figure}

Since we are optimizing the \textit{expected score}, as defined in Eq.~\eqref{eq:exp_score_tot}, and individual evacuation probabilities will almost certainly never be equal to one, we do not expect to recover the best possible score of $342$. Over the parameter ranges $0.5 \leq a \leq 1$ and $3 \leq b \leq 10$, we find that the expected optimal static score, calculated over all $48$ unique evacuation trajectories, is equal to
 \begin{equation}
 \langle S^s \rangle_{opt} = 300.27
 \end{equation}
and 
is achieved at the parameter values $a = 1$, $b = 5.90$. Fig.~\ref{fig:param_space_opt_ind50} shows the parameter space locations of the static optimal strategy, as well as several strategies that give near-optimal expected scores. 

We also solve for the best strategy under the assumption that a player is an optimal Bayesian observer, updating the strategy after each trial to the decision model with the highest posterior probability of producing the maximum score. Because the participants undergo 16 ``test trials'' before the experiment, participants have some experience with the game at the start of the experiment. Hence, these test trials are appended to the 128 main trials in the determination of the Bayesian optimal strategies. We assume a uniform prior over the parameter space $0.5 \leq a \leq 1$ and $3 \leq b \leq 10$.

\begin{figure}[h!]
\includegraphics{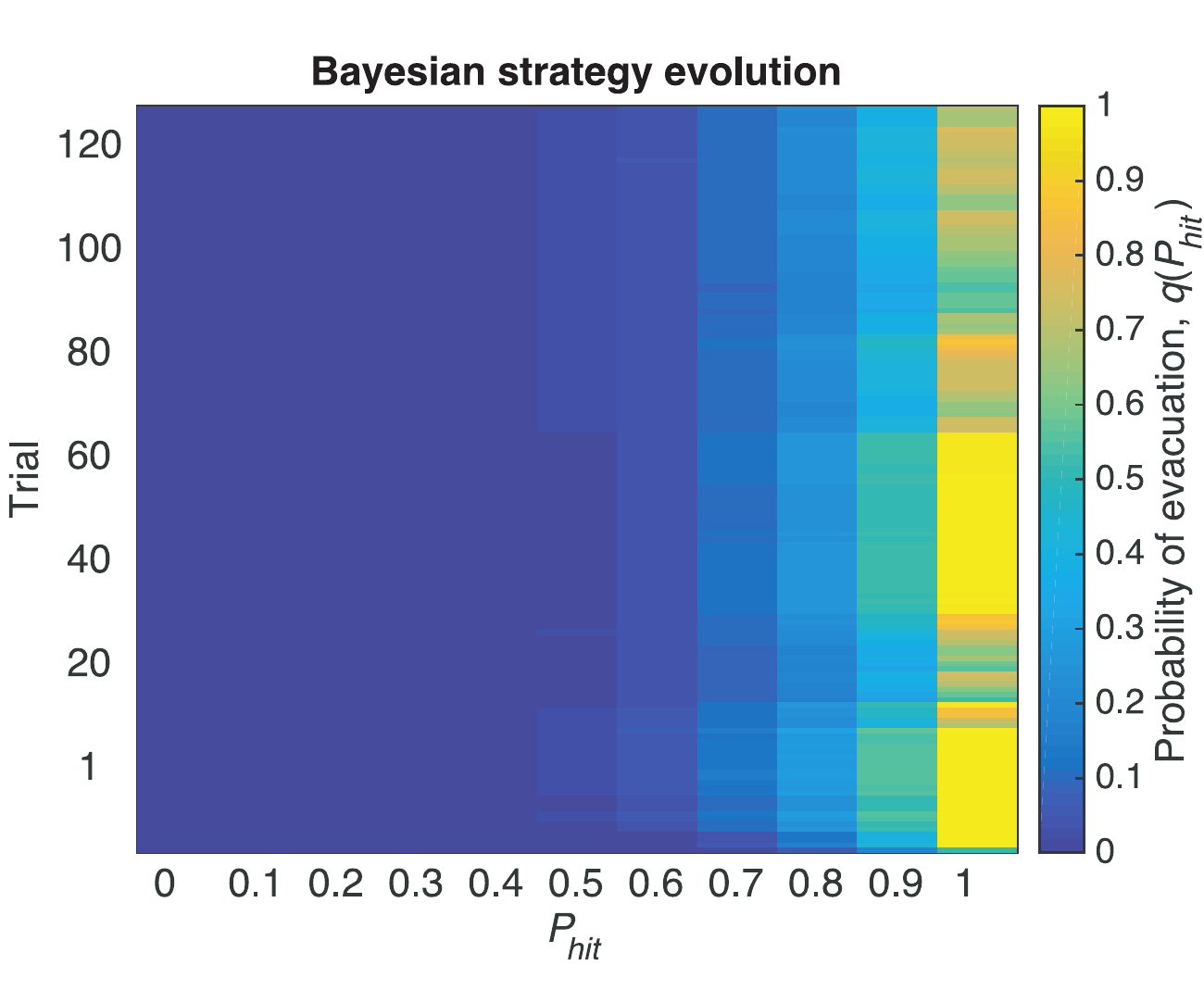}
\caption{{\bf Bayesian optimal strategy evolution.} This plot depicts the evolution of the Bayesian optimal strategy over the course of the experiment, with strategy updates after each trial. After each trial, the color value represents the evacuation probability at each $P_{hit}$ value corresponding to the decision model $q$ with the highest posterior probability of producing the maximum score (under an individual decision protocol). Each strategy update decision takes into account evidence from all previously experienced trials.}
\label{fig:bayes_evol_color}
\end{figure}

\begin{figure}[h!]
\includegraphics{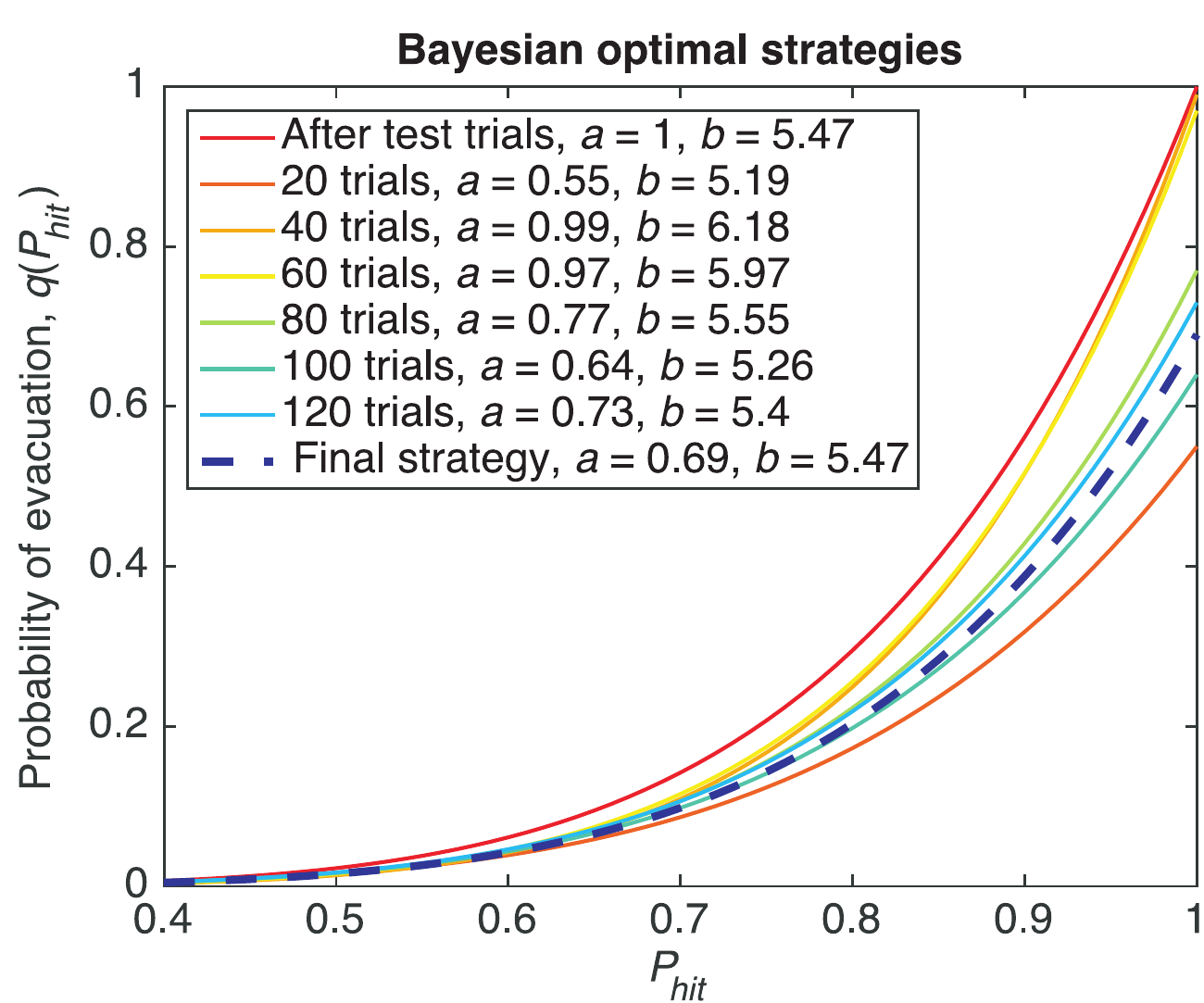}
\caption{{\bf Bayesian optimal strategies.} Unique strategies adopted by the Bayesian optimal player. Warmer colors correspond to earlier strategies, based on evidence from fewer trials. ``Final'' strategy is based on evidence from all $16$ test trials and $128$ main trials presented sequentially.} 
\label{fig:bayes_line}
\end{figure}

Fig.~\ref{fig:bayes_evol_color} shows the time evolution of the \textit{Bayesian optimal} strategy, as updated by the player after each of the $16$ sequential test trials plus $128$ main trials. The colors correspond to the evacuation probability $q(\phit)$. Several example strategies at various points in the experiment are illustrated in Fig.~\ref{fig:bayes_line}.

\begin{figure}[h!]
\includegraphics{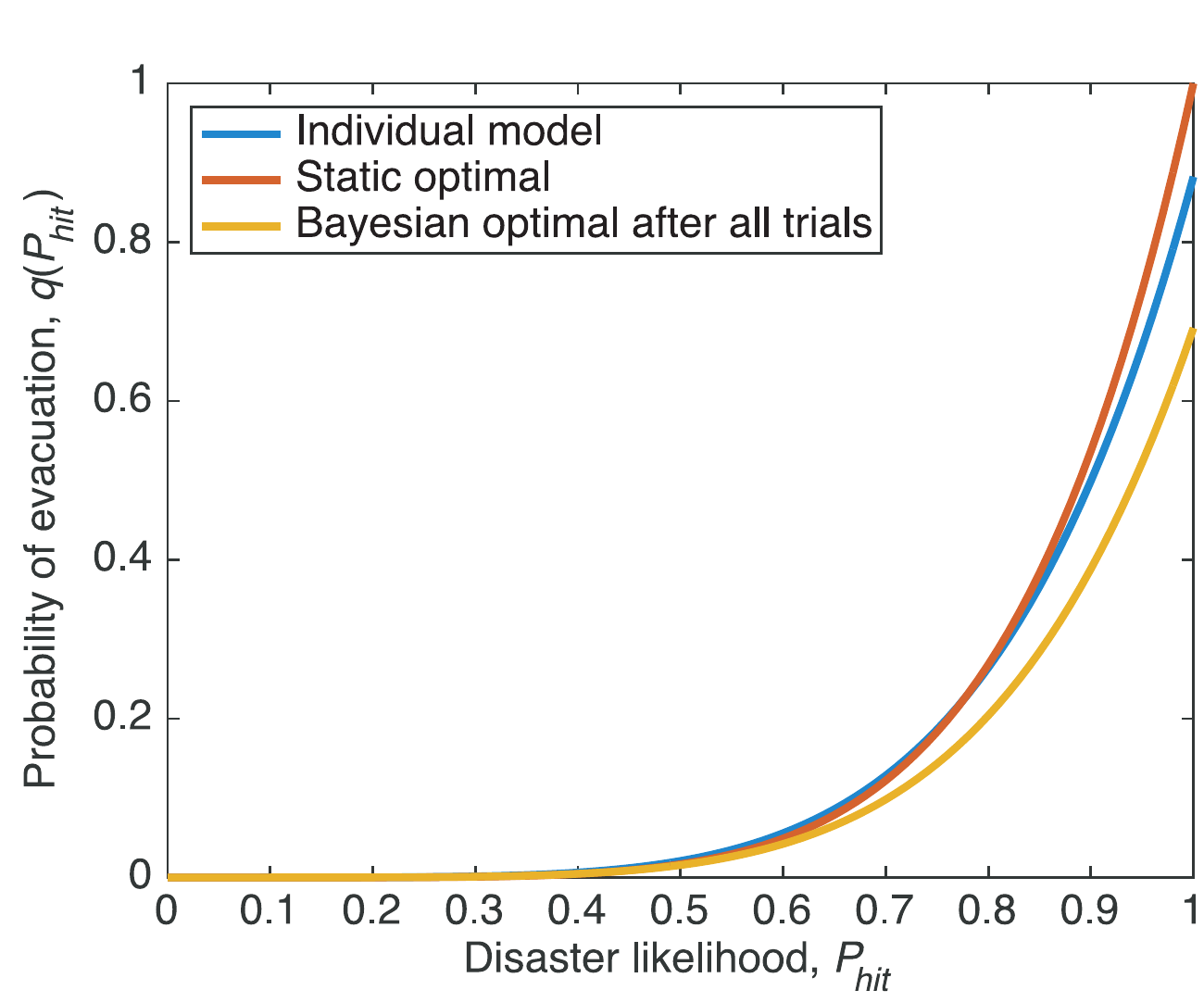}
\caption{{\bf Comparison of optimal and empirical strategies.} The best-fit strategy to the empirical data from all individual trials with 50 shelter spaces is shown in blue (identical to the curve in Fig.~\ref{fig:powermodel}), in comparison to the optimal strategies. The red curve depicts the static optimal strategy, while the yellow curve shows the final Bayesian optimal strategy at the end of the experiment.}
\label{fig:strategies}
\end{figure}

Fig.~\ref{fig:strategies} compares the static optimal and final Bayesian optimal strategies with the original power-law decision model trained on individual 50-space trials. The original model and static optimal strategy are relatively similar, while the Bayesian strategy has a maximum of $0.69$ and is thus more conservative than the other two decision models.

The optimal Bayesian player can achieve a maximum expected score of $802.9$ over the $128$ main trials undergone in sequential order while having trained over the 16 test trials before the start of the first main trial (scores from the test trials are not included). We assume that the player randomly selects a strategy for the trial, thus resulting in an expected score of $9.7$, which is achieved by taking the average values of $a$ and $b$ over uniformly-gridded parameter space.

The maximum possible score after the 128 main trials is 920, while the highest score achieved by a participant was 796. The average score was 763. The expected score of 802 obtained following the evolving Bayesian strategies is only slightly higher than the highest score achieved during the experiment. The static optimal expected score after the 128 trials is 812, higher than the Bayesian score, which would be expected given the ``all-knowing'' assumption of the static optimal strategy compared to the time-dependent revelation of information in the Bayesian case. 

This Bayesian expected score was determined with the assumption of an individual participant unconstrained by limited shelter space or group protocol, but this assumption holds only for Ind50 trials. Thus, we compare the expected scores for the sixteen Ind50 trials, where the Bayesian expected score is calculated using the strategies as updated immediately preceding each of the Ind50 trials in their actual position in the full sequence of the 128 trials. The Bayesian expected score is 112, which is again higher than the actual average score of 91.

In many cases, participants clearly behave sub-optimally. For instance, in many trials, some number of participants evacuate at early time steps, even though there is no competition for shelter space and $\phit$ is relatively low. Furthermore, the trials are not slated to end before $t=20$, yet some participants evacuate before $t=20$. We also observe that in trials immediately following those where $\phit$ increased quickly to 1, resulting in maximum losses for participants who were unable to evacuate in time, a number of participants tend to evacuate very early, regardless of $\phit$ or shelter capacity.

\section*{Discussion}
In this work, we present a model for the collective evacuation behavior observed in a controlled experiment simulating a natural disaster. The experiment setup allows us to isolate and vary several key parameters in order to analyze their influence on decision making in such scenarios. We introduce a decision model that characterizes individual behavior as a function of the disaster likelihood $\phit$ and the initial capacity of the evacuation shelter, with an explicit functional form derived from statistics of our empirical observations. 

We take advantage of the flexibility of the model to quantitatively answer specific questions about how evacuation decisions differ under various conditions. We use the individual model to simulate ``grouped individual'' behavior, quantifying the extent to which decision dynamics change under group constraints. We also determine optimal decision strategies under this model, and compare observed performance with the expected decisions of optimal Bayesian evacuators.

\textbf{Comparison to previous work.} 
The behavioral experiment discussed here is first introduced in Nguyen et al.~\cite{fangqiu}, a work which describes the experiment in detail, discusses statistics of the observed behavior, and then develops a series of incrementally complex artificial neural network models as a data-driven method of predicting the evacuation times of both individuals and groups. These models predict the probability of evacuation on a given time step by considering several factors, including the disaster likelihood ($\phit$) and shelter capacity on the time step in question, the past several values of $\phit$ and shelter capacity, the size and decision protocol of decision-making groups, as well as a personality factor which is determined either from training on the data or from participants' Facebook activity. These models achieve accurate and unbiased predictions, and provide valuable insight into the information on decisions contained in different combinations of experimental factors. However, their reliance on an extensive training data set and use of a large number of input features make it difficult to incorporate their results into large-scale simulations of collective behavior, especially when such efforts lack sufficient quantitative data on specific factors and decisions made in natural disasters. 

In this work, we pursue a distinct but complementary modeling approach. In contrast to the neural network, this model is relatively simple, with the aim of isolating the effects of specific variables and tradeoffs on behavior. However, it provides fairly accurate population-level predictions of evacuation decisions without overfitting, as shown by the results of leave-one-out cross-validation, and it captures generalizable features of collective behavior that can quantitatively inform models on larger scales and in related situations.

The experiment described here and in Nguyen et al.~\cite{fangqiu}, as well as the modeling work in this paper, were inspired by a previous study that laid the groundwork for this type of empirically-driven investigation of evacuation decision making~\cite{sean}. However, a number of modifications are made here to pursue additional questions and scenarios involving social factors. The primary change in the experimental design concerns the addition of trials with forced group consensus decisions; this allows investigation of the behavior of such groups and compare it to individual behavior. We find that the consideration of whether individuals are constrained to make decisions in groups is a crucial aspect when modeling population evacuation behavior, and that the decision protocol used in these groups can drastically alter the resulting dynamics.

Carlson et al.~\cite{sean} modeled evacuation decision likelihood as a function of $\phit$ with a Hill-function decision model, which has a logistic shape with three parameters describing its maximum value, midpoint, and steepness. The resulting decision model is nearly flat for low values of $\phit$, increases sharply for moderate values of $\phit$, and levels off for high values of $\phit$. However, in this work, we find that a Hill function fitted to the data does not level off for high values of $\phit$, and is thus virtually identical to a two-parameter power-law model; therefore, we choose the power-law model in order to parametrize the model with as few parameters as possible. 

In Carlson et al.~\cite{sean}, shelter capacities of 10, 20, 30, 40, and 50 spaces for a total population of 50 were studied, and shelter capacity was found to be linearly related to the midpoint parameter of the Hill function -- i.e., the $\phit$ threshold at which evacuations were observed to rise. In this work, we focus on shelter capacities of 5, 25, and 50 spaces, and find that the observed behavior cannot be captured by a linear dependence of the model on shelter space. In trials with only 5 shelter spaces, competition for shelter space becomes the primary influential factor on decision making, and participants evacuate almost immediately after the start of the trial regardless of $\phit$.

\textbf{Discussion of results.} 
We fit this two-parameter decision model to the statistics of evacuation decisions as a function of $\phit$, as measured in individual decision trials with sufficient shelter space for the entire population, and assess its prediction accuracy for the time-dependent evacuation behavior of the population. Solving the master equation for the full probability distributions of population evacuation outcomes, we show that the total expected number of evacuations predicted by the model tracks well with the observed evacuations (Fig.~\ref{fig:indivtrials}) -- a remarkable level of accuracy for a two-parameter model that does not take the timing of decisions into account, but only the instantaneous $\phit$ value on any given time step. This demonstrates that the perceived $\phit$ value is a driving factor in participants' evacuation decisions. Moreover, these predictions do not significantly change when using leave-one-out cross-validation, indicating that this model is expected to generalize to predict behavior on future, unseen individual trials with similar accuracy.

Despite the accuracy of this model in predicting population-level behavior, it fails to capture certain phenomena. Some discrepancies occur in early time steps: while several participants often evacuate very early -- even in the first time step -- the model predicts lower evacuation numbers than observed in these early time steps because $\phit$ is low. We observe that early evacuations also tend to occur in trials which follow those where $\phit$ rose very quickly to 1. We also observe ``cascading'' behavior in some trials, where a few initial evacuations can lead to a large wave of evacuations. Furthermore, some evacuations occur after $\phit$ has risen steadily over consecutive time steps, even though the value of $\phit$ itself is relatively low. These observations indicate that while the instantaneous information on the disaster likelihood is perhaps the most important factor in determining evacuation decisions, factors such as the timing (i.e., whether a time step is early or late in a trial), social pressure, and trends in the overall movement of $\phit$ over several time steps also influence individual participants' decisions in ways that cannot be captured by focusing only on $\phit$.

At the population level, the assumption that participants are independent and identically distributed leads to accurate prediction of individual trials. However, we show with ``grouped individual'' simulations that that accurate characterization of individual behavior alone is insufficient to fully capture the behavior of groups. Although this is likely due to individuals adjusting their strategies when acting in groups, individual differences may also play a role. Differences between the decision-making tendencies of individuals may be largely averaged out in the model of individual decision trials, but still give a reasonable approximation to the population-level behavior. However, individual differences may have a larger influence in forced group decision scenarios where one individual's decision controls the action of the entire group -- especially in scenarios in which influence is skewed toward participants with extreme strategies. For example, first-to-go group protocols allow single decisions by exceptionally early evacuators to impose the same dynamics on the four or twenty-four other participants in their groups, regardless of the personal preferences of those other participants, which might have counterbalanced the early evacuation tendencies of their peers in a individual or majority-vote decision scenario. Because the ``grouped individual'' simulations maintain the assumption of independent and identically distributed evacuation statistics for all participants, they would not capture the effect of a small number of individuals who were particularly skewed toward early evacuation. Nguyen et al.~\cite{fangqiu} uses an individualized personality factor as an extra input feature to the data-driven decision model, which allows more accurate prediction of evacuation times in both individual and group trials; future modeling work will expand the decision model in the present work to include individual difference factors.

This work compares three different decision protocols that may alter the evacuation success rate of groups in the disaster scenario. There is little distinction in success rate across protocols for group-5 trials, but majority-vote generally results in the highest scores for group-25 trials, as discussed in Nguyen et al.~\cite{fangqiu}. Majority-vote is a very common decision protocol observed in both hunter-gatherer tribal societies and modern organizations~\cite{majority}. It has been observed that when groups are not constrained to act according to any protocol and receive information uniformly, they tend to make decisions according to a majority vote~\cite{eckstein}. Here, to isolate the effect of decision mechanisms and specific social information (member ranks and decisions) on behavior, the decision protocols were fixed for each trial and known to all participants, although group members could not directly communicate and instead were required to make decisions based on the ranks and actions of others. 

\textbf{Relation to real-world scenarios.} 
In real-world disaster scenarios, information is often limited and uncertain. Information regarding the progression of a disaster is usually reported as a likelihood or probability. The presentation of probabilistic information, e.g., whether it is displayed as discrete values or a continuous distribution, has been shown to influence decision making based on this information~\cite{edwards}. We deliberately chose to represent disaster progression as a discretized value labeled as a likelihood rather than a probability, and to round the value to reflect the limited amount of information available in a disaster scenario. In real-world scenarios, $\phit$ values can be calculated to inform public policy decisions and evacuation orders. For example, hurricane strike probabilities can be calculated from wind velocities and other factors for up to 120 hours into the future~\cite{regnier}. Ultimately, the interpretation of warning information influences the assessment of potential risks and subsequent action or inaction, and understanding these responses can shape the effectiveness of warning messages and evacuation orders~\cite{evachouse}.

In addition, our choice to study the dynamics of decisions by group consensus reflects the reality of many real-world evacuation scenarios. Often, evacuation decisions are made within a household, whose members are often bound to the same decision due to transportation and other needs. These groups may use a variety of strategies to make decisions, depending on the ages, relationships, and independence of their members. This work demonstrates that group composition and group decision mechanisms are important influences on models of population-level behavior during a natural disaster.

The communities studied in this instance are arranged in groups of various sizes, a very specific special case of a ``networked'' community as it is traditionally studied. Groups of small sizes play an important role in evacuation decisions that are often centered around households; however, much of the dynamics in real disaster scenarios also plays out atop networked systems at different scales, especially those supporting communication (mobile and broadcast networks, social media, etc.) and transportation (public transit, evacuation routes, etc.). These networks place important constraints on the flow of information that influences evacuation decisions and the successful movement required for evacuation. Future experiments will incorporate such networks explicitly, in order to examine their interactions with other decision-making behaviors, and to derive best practices for interacting with these networks in light of the influence of groups on the collective dynamics that these networks support.

In real-world disaster scenarios and other high-pressure situations, asymmtries in human decision making and risk aversion are common and can significantly influence outcomes~\cite{sean,chib2012,bach2012,risk1,risk2}. Complex, large-scale models of population-level behavior often model human decisions as random, optimal, or a simple threshold function of an opinion state variable~\cite{dani,stir12,stir13,helbing1,helbing2}. In contrast, our empirically based model is able to isolate and quantify asymmetries in individual decision making that impact population behavior. For example, previous work has shown that incorporating the asymmetric power-law response of individual evacuation probability as a function of $\phit$ substantially improves predictions and fits to the observed behavior, when compared to modeling this response as a simpler linear or threshold function~\cite{imtiaz}. Our modeling approach captures this aspect of the dynamics succinctly, which gives insight into similar behavior in related situations and can feasibly be incorporated into a larger-scale simulation.

\textbf{Limitations and future work.}
Future studies will investigate the effect of heterogeneous information broadcast on collective decision making. Heterogeneous updates can lead to more disparate success rates across individuals, such that the accuracy of group decisions can significantly increase if greater weight is assigned to the most successful individuals' decisions~\cite{decisionsurvey}. Furthermore, when individuals are free to decide according to any protocol, if information broadcast is heterogeneous, majority-vote decisions are often swayed by one or more group members who hold an opposing opinion with high confidence~\cite{eckstein}. Hence, emergent group leadership may become more apparent with heterogeneous update, such that those who receive updates at the most frequent rates may hold greater influence over their group's decisions. 

As the participants in the experiment were primarily male students under the age of 30, future studies can involve experiments with a more diverse set of participants. Evacuation decisions are influenced by age and socioeconomic status, and risk perception varies with age, cultural background, and experience~\cite{evachouse,gladwinpeacock}. In real-world disaster scenarios, decisions can also be influenced by geographic location, household or group size, and the presence of children, elderly household members, or pets~\cite{evachouse,huang2011,heath2001human}. 

While this controlled experimental setting remains somewhat removed from a real-world disaster scenario, our findings provide important insights into individual and group decision making which can inform the design of protocols for action in the case of natural or man-made disasters. Quantifying the optimal upper-bound of decision making and where human behavior deviates from the optimum serves to caution against policies based on collective behavior models and agent-based simulations which treat participants as perfectly optimal or completely random. Furthermore, as individual behavior does not fully predict group behavior, the effect of social information is crucial to the development of public policy for collective action under stress or threat.

\section*{Supporting Information}
\beginsupplement

\begin{figure*}[ht]
\includegraphics[width=\linewidth]{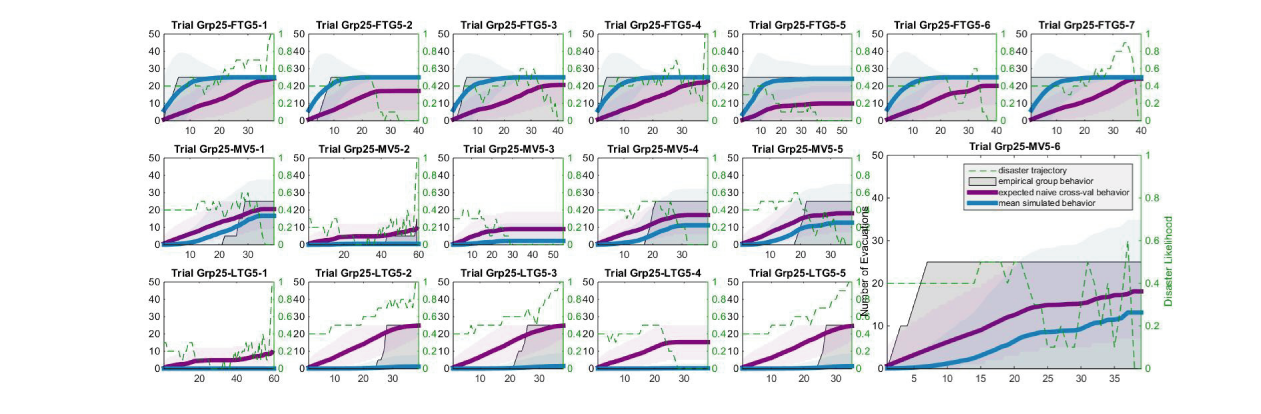}
\caption{\textbf{Individual versus group decision strategies under competition in group-5 trials.} Empirically observed cumulative evacuations (gray shaded area) on group-5 trials, with initial shelter space of 25. Na\"ive cross-validation predictions are shown in solid purple, with 99.7\% confidence interval in magenta, and grouped-individual simulation results are shown in solid blue (average over 1000 simulations), with confidence interval in light blue. All three decision protocols -- first-to-go (FTG), majority-vote (MV), and last-to-go (LTG) -- are shown. Selected example trials from all three protocols are highlighted in Figure \ref{fig:ind_grp_trials25} in the main manuscript.}
\label{fig:alltrials_25_5}
\end{figure*}

\begin{figure*}[ht]
\includegraphics[width=\linewidth]{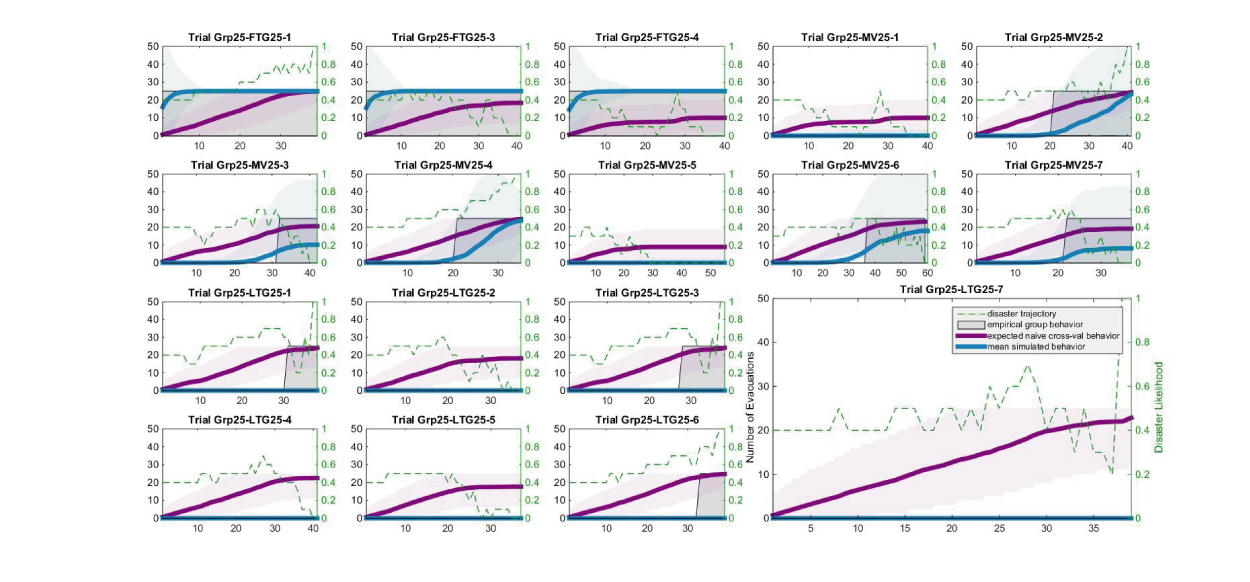}
\caption{\textbf{Individual versus group decision strategies under competition in group-25 trials.} Empirically observed cumulative evacuations (gray shaded area) on group-25 trials, with initial shelter space of 25. Na\"ive cross-validation predictions are shown in solid purple, with 99.7\% confidence interval in magenta, and grouped-individual simulation results are shown in solid blue (average over 1000 simulations), with confidence interval in light blue. All three decision protocols -- first-to-go (FTG), majority-vote (MV), and last-to-go (LTG) -- are shown. Selected example trials from all three protocols are highlighted in Figure \ref{fig:ind_grp_trials25} in the main manuscript.}
\label{fig:alltrials_25_25}
\end{figure*}

\section*{Acknowledgments}
This work was supported by the David and Lucile Packard Foundation (www.packard.org) and the Institute for Collaborative Biotechnologies (www.icb.ucsb.edu) through contract no. W911NF-09-D-0001 from the U.S. Army Research Office (http://www.aro.army.mil/). KJS was supported by the National Science Foundation Graduate Research Fellowship Program (www.nsfgrfp.org) through grant no. DGE-1144085. KJS and IA were additionally supported by the Susan and Bruce Worster Fellowship for physics research at UC Santa Barbara. The funders had no role in study design, data collection and analysis, decision to publish, or preparation of the manuscript. The content of the information does not necessarily reflect the position or the policy of the Government, and no official endorsement should be inferred.




\end{document}